\documentclass{cimsart}                    

\received{Month 199X}       
\yearofpublication{199X}                \volume{000}
\startingpage{1}                        \cccline{00}

\authorheadline{P.J.~Forrester and N.S.~Witte}
\titleheadline{Application of the $\tau$-function theory}

\usepackage{amsmath}                         
\usepackage{amssymb}                         

\newtheorem{lemma}[theorem]{Lemma}
\newtheorem{cor}[theorem]{Corollary}
\newtheorem{prop}[theorem]{Proposition}

\renewcommand{\theequation}{\thesection.\arabic{equation}}
\newcommand{\zz}{\mathbb Z}
\newcommand{\+}{\!+\!}
\newcommand{\m}{\!-\!}
\newcommand{\uU}{\overline{U}}
\newcommand{\uuU}{\overline{\overline{U}}}
\newcommand{\dU}{\underline{U}}
\newcommand{\tU}{\overrightarrow{U}}
\newcommand{\ttU}{\overrightarrow{\overrightarrow{U}}}
\newcommand{\bU}{\overleftarrow{U}}
\newcommand{\PII}{${\rm P}_{\rm II}\;$}
\newcommand{\PIII}{${\rm P}_{\rm III}\;$}
\newcommand{\PIIIa}{${\rm P}_{\rm III^{\prime}}\;$}
\newcommand{\PIV}{${\rm P}_{\rm IV}\;$}
\newcommand{\PV}{${\rm P}_{\rm V}\;$}

\begin{document}                        



\title{
Application of the $\tau$-function theory of Painlev\'e equations to
random matrices: \\ \PV, \PIII, the LUE, JUE and CUE}



\author{P.J.~Forrester \\            
\affil 
Department of Mathematics and Statistics,\\
\it University of Melbourne, Victoria 3010, Australia \\  
\\ AND \\
\\ N.S.~Witte \\               
\affil
Department of Mathematics and Statistics, and School of Physics,\\ 
\it University of Melbourne, Victoria 3010, Australia \\  
}                                       




\maketitle   


\begin{abstract}
With $\langle \cdot
\rangle$ denoting an average with respect to the  eigenvalue PDF for
the Laguerre unitary ensemble, the object of our study is 
$ \tilde{E}_N(I;a,\mu) := \langle \prod_{l=1}^N \chi_{(0,\infty)\backslash I}^{(l)}
(\lambda - \lambda_l)^\mu \rangle $
for $I = (0,s)$ and $I = (s,\infty)$, where $\chi_I^{(l)} = 1$ for
$ \lambda_l \in I$ and $\chi_I^{(l)} = 0$ otherwise. Using Okamoto's
development of the theory of the Painlev\'e V equation, it is shown
that $\tilde{E}_N(I;a,\mu)$ is a $\tau$-function associated with the
Hamiltonian therein, and so can be characterised as the solution of
a certain second order second degree differential equation, or in terms of
the solution of certain difference equations. The cases
$\mu = 0$ and $\mu = 2$ are of particular interest as they correspond to the
cumulative distribution and density function respectively for the
smallest and largest eigenvalue. In the case $I = (s,\infty)$,
$\tilde{E}_N(I;a,\mu)$ is simply related to an average in the Jacobi
unitary ensemble, and this in turn is simply related to certain averages
over the  orthogonal group, the unitary symplectic group and the
circular unitary ensemble. The latter integrals are of interest for their
combinatorial content. Also considered are the hard edge and soft edge scaled
limits of $\tilde{E}_N(I;a,\mu)$. In particular, in the hard edge scaled
limit it is shown that the limiting quantity $E^{\rm hard}((0,s);a,\mu)$
can be evaluated as a $\tau$-function associated with the Hamiltonian in
Okamoto's theory of the Painlev\'e III equation. 
\end{abstract}



\tableofcontents



\section{Introduction and summary}
In a previous paper \cite{FW00} the quantities
\begin{equation}\label{1.a}
  \tilde{E}_N(\lambda;a) := \Big \langle
  \prod_{l=1}^N \chi_{(-\infty, \lambda]}^{(l)} (\lambda - \lambda_l)^a
  \Big \rangle_{\rm GUE}, \quad
  \chi_{(-\infty, \lambda]}^{(l)} =
       \begin{cases}
              1 & \lambda_l \in (-\infty, \lambda], \\
              0 & \text{otherwise}
       \end{cases}
\end{equation}
and
\begin{equation}\label{1.b}
  F_N(\lambda;a) := \Big \langle \prod_{l=1}^N(\lambda - \lambda_l)^a
  \Big \rangle_{\rm GUE},
\end{equation}
where the averages are with respect to the joint eigenvalue distribution
of the Gaussian unitary ensemble (GUE), were shown to be equal to the
$\tau$-functions occurring in Okamoto's theory \cite{Ok86} of the
Painlev\'e IV equation. It was noted in \cite{FW00} that we expect the
analogous quantities for the Laguerre unitary ensemble (LUE) to be
expressible in terms of the $\tau$-functions occurring in 
Okamoto's theory \cite{Ok87} of the Painlev\'e V equation. It is the
purpose of this article to verify this statement by giving the details
of the correspondences between the multi-dimensional integrals defining
the analogues of $\tilde{E}_N(\lambda;a)$ and $F_N(\lambda;a)$ for the
LUE, and the $\tau$-functions from \cite{Ok87}.

Let us first recall the definition of the LUE. Let $X$ be a $n \times
N$ $(n \ge N)$ Gaussian random matrix of complex elements $z_{jk}$,
with each element independent and distributed according to the Gaussian
density ${1 \over \pi} e^{-|z_{jk}|^2}$ so that the joint density of
$X$ is proportional to 
\begin{equation}\label{1.1}
  e^{-{\rm Tr} X^\dagger X}.
\end{equation}
Denote by $A$ 
the non-negative matrix $X^\dagger X$. Because (\ref{1.1}) is unchanged
by the replacement $X \mapsto UXV$ for $U$ a $n \times n$ unitary
matrix and $V$ a $N \times N$ unitary matrix, the ensemble of matrices
$A$ is said to have a unitary symmetry. The probability density
function (PDF) for the eigenvalues of $A$ is given by
\begin{equation}\label{1.2}
  {1 \over C} \prod_{l=1}^N \lambda_l^a e^{-\lambda_l}
  \prod_{1 \le j < k \le N} (\lambda_k - \lambda_j)^2, \qquad
  \lambda_l >0,
\end{equation}
where $C$ denotes the normalization and $ a=n-N, \, n \ge N $. 
(Throughout, unless otherwise stated, the symbol $C$ will be
used to denote {\it some} constant i.e.~a quantity independent of the
primary variables of the equation.) Because $\lambda^ae^{-\lambda}$ is the
weight function occurring in the theory of the Laguerre orthogonal
polynomials, and the ensemble of matrices $A$ has the aforementioned
unitary symmetry, the eigenvalue PDF (\ref{1.2}) is said to define the
LUE.

Analogous to $\tilde{E}_N(\lambda;a)$ specified by (\ref{1.a}) for the
GUE, we introduce the quantities 
\begin{align}
  \tilde{E}_N((0,s);a,\mu) & := \Big \langle
  \prod_{l=1}^N \chi_{(s, \infty)}^{(l)} (\lambda_l-s)^\mu
  \Big \rangle_{\rm LUE} \label{2.b}\\
  \tilde{E}_N((s,\infty);a,\mu) & := \Big \langle
  \prod_{l=1}^N \chi_{(0, s)}^{(l)} (s - \lambda_l)^\mu
  \Big \rangle_{\rm LUE} \label{2.a}
\end{align}
where the averages are with respect to (\ref{1.2}) (the parameter
$a$ in (\ref{1.a}) has been denoted $\mu$ in (\ref{2.b}), (\ref{2.a}) to 
avoid confusion with the parameter $a$ in (\ref{1.2})). Explicitly
\begin{multline}\label{2.d}
  \tilde{E}_N((0,s);a,\mu) \\
  =  {1 \over C}
  \prod^{N}_{l=1}\int_s^\infty d\lambda_{l} \, 
                 \lambda_{l}^a e^{-\lambda_{l}}(\lambda_{l}-s)^\mu
  \prod_{1 \le j < k \le N}(\lambda_k - \lambda_j)^2 \\
  =  {e^{-Ns} \over C}
  \prod^{N}_{l=1}\int_0^\infty d\lambda_{l} \, 
                 \lambda_{l}^\mu (\lambda_{l} + s)^a e^{-\lambda_{l}}
  \prod_{1 \le j < k \le N} (\lambda_k - \lambda_j)^2 \\
  =  {e^{-Ns} \over C} s^{(a+\mu)N + N^2}
  \prod^{N}_{l=1}\int_0^\infty d\lambda_{l} \, 
                 \lambda_{l}^\mu (\lambda_{l} + 1)^a e^{-s\lambda_{l}}
  \prod_{1 \le j < k \le N}(\lambda_k - \lambda_j)^2 
\end{multline}
\begin{multline}\label{2.c}
  \tilde{E}_N((s,\infty);a,\mu) \\
  =  {1 \over C}
  \prod^{N}_{l=1}\int_0^s d\lambda_{l} \, 
                 \lambda_{l}^a e^{-\lambda_{l}}(s-\lambda_{l})^\mu 
  \prod_{1 \le j < k \le N}(\lambda_k - \lambda_j)^2 \\
  = {1 \over C}s^{(a+\mu)N + N^2}
  \prod^{N}_{l=1}\int_0^1 d\lambda_{l} \, 
                 \lambda_{l}^a (1-\lambda_{l})^\mu e^{-s\lambda_{l}}
  \prod_{1 \le j < k \le N}(\lambda_k - \lambda_j)^2 .
\end{multline}

In the case $\mu = 0$ the first integrals in (\ref{2.d}) and (\ref{2.c})
are the definitions of the probability that there are no eigenvalues in
the intervals $(0,s)$ and $(s,\infty)$ respectively of the Laguerre
unitary ensemble. The case $\mu = 2$ also has significance in this
context. To see this, first note from the definitions (using the first integral
representation in each case) that
\begin{align}
  {d \over ds} \tilde{E}_{N+1}((0,s);a,0) 
  & \propto s^a e^{-s} \tilde{E}_{N}((0,s);a,2) \label{u1u} \\
  {d \over ds} \tilde{E}_{N+1}((s,\infty);a,0) 
  & \propto s^a e^{-s} \tilde{E}_{N}((s,\infty);a,2). \label{u2}
\end{align}
On the other hand, with $p_{\rm min}(s;a)$ and $p_{\rm max}(s;a)$
denoting the distribution of the smallest and largest eigenvalue
respectively in the $N \times N$ LUE, we have
\begin{align}
  p_{\rm min}(s;a) 
  & = -{d \over ds} \tilde{E}_{N}((0,s);a,0) \label{R0} \\
  p_{\rm max}(s;a) 
  & = {d \over ds} \tilde{E}_{N}((s,\infty);a,0). \label{R1}
\end{align}
Under the replacement $N \mapsto N + 1$, $p_{\rm min}(s;a)$ and
$p_{\rm max}(s;a)$ are determined by $\tilde{E}_{N}((0,s);a,2)$
and $\tilde{E}_{N}((s,\infty);a,2)$ respectively.

From the second formula in (\ref{2.d}), we see that with
\begin{equation}\label{2.d1}
  F_N(s;a,\mu) := \Big \langle \prod_{l=1}^N(\lambda_l - s)^\mu \Big 
  \rangle_{\rm LUE}
\end{equation}
we have
\begin{equation}\label{13}
  F_N(s;a,\mu) = \Big ( e^{Ns} \tilde{E}_N((0,s);\mu,a) \Big )
  \Big |_{s \mapsto - s}
\end{equation}
(notice the dual role played by $\mu$ and $a$ on the different sides of
(\ref{13}))
so there is no need to consider $F_N$ separately. Note that with
$\mu = 2$, (\ref{2.d1}) multiplied by $s^a e^{-s}$ is
proportional to the definition of the density in the LUE with $N \mapsto N+1$.
We also remark that for $ \mu $ non-integer the definition (\ref{2.d1}) must 
be complemented by a definite choice of branch of the function 
$ (\lambda_l - s)^\mu $. But this case does not appear in our random matrix 
applications so we will not address the issue further.

Some quantities generalizing (\ref{2.d}) and (\ref{2.c}), and which
include (\ref{2.d1}) are 
\begin{multline}\label{1.14a}
  \tilde{E}_N((0,s);a,\mu;\xi) \\
  := {1 \over C}
  \prod^{N}_{l=1} \Big( \int_0^\infty - \xi \int_0^s \Big) d\lambda_{l} \,
  \lambda_{l}^a e^{-\lambda_{l}} (\lambda_{l} - s)^\mu
  \prod_{1 \le j < k \le N} (\lambda_k - \lambda_j)^2
\end{multline}
\begin{multline}\label{1.14b}
  \tilde{E}_N((s,\infty);a,\mu;\xi) \\
  := {1 \over C}
  \prod^{N}_{l=1} \Big( \int_0^{\infty} - \xi \int_s^{\infty} \Big) d\lambda_{l} \,
  \lambda_{l}^a e^{-\lambda_{l}} (s-\lambda_{l})^\mu
  \prod_{1 \le j < k \le N} (\lambda_k - \lambda_j)^2.
\end{multline}
Thus
\begin{align*}
  \tilde{E}_N((0,s);a,\mu;0) & = F_N(s;a,\mu) \\
  \tilde{E}_N((0,s);a,\mu;1) & = \tilde{E}_N((0,s);a,\mu) \\
  \tilde{E}_N((s,\infty);a,\mu;1) & = \tilde{E}_N((s,\infty);a,\mu). 
\end{align*}
Note that only one of the quantities (\ref{1.14a}), (\ref{1.14b}) is
independent since
\begin{equation*}
  \tilde{E}_N((s,\infty);a,\mu;\xi) = e^{-\pi iN\mu}
  \Big ( {1 \over (1 - \lambda)^N}
  \tilde{E}_N((0,s);a,\mu;\lambda) \Big ) \Big |_{\lambda = \xi/(\xi - 1)}.
\end{equation*}
Their interest stems from the facts that
\begin{align*}
  {(-1)^n \over n!} {\partial^n \over \partial \xi^n}
  \tilde{E}_N((0,s);a,0,\xi) \Big |_{\xi = 1} & = 
  E_N(n;(0,s)) \\
  {(-1)^n \over n!} {\partial^n \over \partial \xi^n}
  \tilde{E}_N((s,\infty);a,0,\xi) \Big |_{\xi = 1} & =
  E_N(n;(s,\infty))
\end{align*}
where $E_N(n;I)$ denotes the probability that in the LUE the interval $I$
contains precisely $n$ eigenvalues.

The second integral in (\ref{2.c}) is of interest for its relevance
to the Jacobi unitary ensemble, which is specified by the eigenvalue
PDF
\begin{equation}\label{J1}
  {1 \over C} \prod_{l=1}^N\lambda_l^a(1 - \lambda_l)^b
  \prod_{1 \le j < k \le N}(\lambda_k - \lambda_j)^2, \qquad
  0 < \lambda_l < 1.
\end{equation}
This ensemble is realised by matrices of the form $A(A+B)^{-1}$,
where $A=X^\dagger X$, $B=Y^\dagger Y$, for $X$ ($Y$) an
$n_1 \times N$ ($n_2 \times N$) complex Gaussian random matrix with
joint density (\ref{1.1}). The parameters $a$ and $b$ are then specified
by $a=n_1 - N$, $b=n_2 - N$ (c.f.~the value of $a$ in (\ref{1.2})).
We see from (\ref{J1}) that
\begin{equation}\label{J1'}
  {1 \over C} \prod^{N}_{l=1}
  \int_0^1 d\lambda_{l}\, \lambda_{l}^a(1-\lambda_{l})^b e^{s\lambda_{l}}
  \prod_{1 \le j < k \le N}(\lambda_k - \lambda_j)^2
  = \Big \langle e^{s \sum_{j=1}^N \lambda_j} \Big \rangle_{\rm JUE},
\end{equation}
so substituting in (\ref{2.c}) and compensating for the different
normalizations in (\ref{2.c}) and (\ref{J1}) shows
\begin{equation}\label{16}
  \tilde{E}_N((s,\infty);a,\mu) = {J_N(a,\mu) \over I_N(a)}
  s^{(a+\mu)N + N^2}
  \Big \langle e^{-s \sum_{j=1}^N \lambda_j} \Big \rangle_{\rm JUE}
  \Big |_{b \mapsto \mu}
\end{equation}
where
\begin{align}\label{2.h1}
  I_N(a) & := 
  \prod^{N}_{l=1}\int_0^\infty d\lambda_{l} \, \lambda_{l}^a e^{-\lambda_{l}}
  \prod_{1 \le j < k \le N} (\lambda_k - \lambda_j)^2 \\
  J_N(a,\mu) & := 
  \prod^{N}_{l=1}\int_0^{1}  d\lambda_{l} \, \lambda_{l}^a (1 - \lambda_{l})^\mu
  \prod_{1 \le j < k \le N} (\lambda_k - \lambda_j)^2. \label{2.h1a}
\end{align}
As is well known, the integrals $I_N(a)$ and $J_N(a,\mu)$
can be evaluated in the form
$N! \prod_{j=0}^{N-1} c_j^2$, where $c_j$ is the normalization of the
monic orthogonal polynomial of degree $j$ associated with the weight functions
$\lambda^a e^{-\lambda}$ and $\lambda^a (1 - \lambda)^\mu$ respectively.

One important feature of the JUE is that with the change of variables
\begin{equation}\label{J2}
  \lambda_j = {1 \over 2} (\cos \theta_j + 1),
\end{equation}
the PDF (\ref{J1}) assumes a trigonometric form, which for appropriate
$(N,a,b)$ coincides with the PDF for the independent eigenvalues
$e^{i \theta_j}$ of random orthogonal and random unitary symplectic
matrices. In the case of orthogonal matrices, one must distinguish the
two classes $O_N^+$ and $O_N^-$ according to the determinant equalling
$+1$ or $-1$ respectively. The cases of $N$ even and $N$ odd must
also be distinguished. For $N$ odd all but one eigenvalue come in
complex conjugate pairs $e^{\pm i \theta_j}$, with the remaining
eigenvalue equalling $+1$ for $O_N^+$ and $-1$ for $O_N^-$. For
$N$ even, all eigenvalues of $O_N^+$ come in complex conjugate pairs,
while for $O_N^-$ all but two eigenvalues come in complex conjugate pairs,
with the remaining two equalling $\pm 1$. Let us replace $N$ in (\ref{J1})
by $N^*$ and make the change of variables (\ref{J2}). Then the PDF for the
independent eigenvalues of an ensemble of random orthogonal matrices
is (see e.g.~\cite{Fo01}) of the form 
(\ref{J1}) with
\begin{equation}\label{17'}
  (N^*,a,b) = \left \{ \begin{array}{ll}
  (N/2,-1/2,-1/2) & {\rm for \, matrices \, in}
  \, O_N^+, \, N \, {\rm even} \\
  ((N-1)/2,-1/2,1/2) & {\rm for \, matrices \, in}
  \, O_N^+, \, N \, {\rm odd} \\
  ((N-1)/2,1/2,-1/2) & {\rm for \, matrices \, in}
  \, O_N^-, \, N \, {\rm odd} \\
  (N/2-1,1/2,1/2) & {\rm for \, matrices \, in}
  \, O_N^-, \, N \, {\rm even}
\end{array} \right.
\end{equation}

Matrices in the group $USp(N)$ are equivalent to $2N \times 2N$ unitary
matrices in which each $2 \times 2$ block has a real quaternion structure.
The eigenvalues come in complex conjugate pairs $e^{\pm i \theta_j}$,
and the PDF of the independent elements is of the form (\ref{J1})
with the change of variables (\ref{J2}) and
\begin{equation}\label{J4}
  (N^*,a,b) = (N,1/2,1/2).
\end{equation}

It follows from the above revision that
\begin{equation}\label{J5}
  \Big \langle e^{s {\rm Tr}(U)} \Big \rangle_{U \in G}
\end{equation}
for $G = O_N^+,O_N^-$ or $USp(N)$ is a special case of the more general
average (\ref{J1'}). 
Explicitly,
\begin{equation}\label{n1}
  \Big \langle e^{s {\rm Tr}(U)} \Big \rangle_{U \in G} =
  e^{-2sN^*} e^{\chi_{N^*} s}
  \Big \langle e^{4s \sum_{j=1}^{N^*} \lambda_j}  \Big \rangle_{\rm JUE}
\end{equation}
where on the RHS the dimension of the JUE is at first $N^*$, then the
parameters $(N^*,a,b)$ are specified as in (\ref{17'}) or (\ref{J4}). 
Also, $\chi_{N^*} = 0$ for $G = USp(N)$ and $O_N^+, O_N^-$ ($N$ even),
while $\chi_{N^*} = \pm 1$ for $G=O_N^+, O_N^-$ ($N$ odd) respectively.
The averages (\ref{J5})
have independent importance due to
their occurrence as generating functions for certain
combinatorial problems. Let us quote one example. Set 
$f_{nl}^{(\rm inv)}$ equal to the number of fixed point free involutions
of $\{1,2,\dots,2n\}$ constrained so that the length of the maximum decreasing
subsequence is less than or equal to $2l$, and introduce the
generating function
\begin{equation}
  P_l(t) := e^{-t^2/2} \sum_{n=0}^\infty
  {t^{2n} \over 2^n n!} {f_{nl}^{(\rm inv)} \over (2n-1)!!}.
\end{equation}
According to a result  of Rains \cite{Ra98} (see also \cite{BF00}) we have
\begin{equation}\label{21}
  P_l(t) = e^{-t^2/2} \Big \langle e^{t {\rm Tr}(U)} \Big \rangle_{U \in 
  USp(l)}.
\end{equation}
A list of similar results is summarised in \cite{AV99b}.

The second integral in (\ref{2.c}) can be written in a trigonometric form
distinct from that which results from the substitution (\ref{J2}).
For this purpose we recall the integral identity \cite{Fo95b}
\begin{multline}\label{1.26'}
  \prod^{N}_{l=1}\int_0^{1} dt_{l}\, t_{l}^{\epsilon-1}\, f(t_1,\dots,t_N) \\ 
  = \Big ( {\pi \over \sin \pi \epsilon} \Big )^N
    \prod^{N}_{l=1}\int_{-1/2}^{1/2}dx_{l} \, e^{2 \pi i \epsilon x_{l}}\,
    f(-e^{2 \pi i x_1}, \dots , -e^{2 \pi i x_N})
\end{multline}
valid for $f$ a Laurent polynomial. Applying (\ref{1.26'}) to
(\ref{2.h1a})  gives the trigonometric integral 
\begin{multline}\label{mn}
  M_N(a',b') := 
    \prod^{N}_{l=1}\int_{-1/2}^{1/2}dx_{l} \, e^{\pi i x_{l}(a'-b')}
    |1 + e^{2 \pi i x_{l}}|^{a'+b'} \\
  \quad \times
  \prod_{1 \le j < k \le N} |e^{2 \pi i x_k} - e^{2 \pi i x_j} |^2.
\end{multline}
Moreover, applying (\ref{1.26'}) to the LHS of (\ref{J1'}) 
and using Carlson's theorem it follows that
\begin{multline}\label{k1}
  \Big \langle e^{s \sum_{j=1}^N \lambda_j} \Big \rangle_{\rm JUE} 
  \Big |_{b \mapsto \mu} \\
  = {M_N(0,0) \over M_N(a',b')}
  \Big \langle \prod_{l=1}^N e^{\pi i x_l(a'-b')} |1 +
  e^{2 \pi i x_l}|^{a'+b'} e^{s e^{2 \pi i x_l}}  \Big \rangle_{\rm CUE},
\end{multline}
where $a'=N+a+\mu$, $b'=-(N+a)$ and the CUE (circular unitary ensemble)
average is over the
eigenvalue PDF
\begin{equation}\label{1.29}
  {1 \over C} \prod_{1 \le j < k \le N}
  | e^{2 \pi i x_k} - e^{2 \pi i x_j} |^2, \qquad
  -1/2 \le x_l \le 1/2.
\end{equation}
Recalling (\ref{16}) then gives 
\begin{multline}\label{k2}
  \Big\langle \prod_{l=1}^N e^{\pi i x_l(a'-b')} |1 +
  e^{2\pi i x_l}|^{a'+b'} e^{-s e^{2\pi i x_l}}  \Big\rangle_{\rm CUE} \\
  = {M_N(a',b') I_N(a) \over M_N(0,0) J_N(a,\mu)}
  s^{-(a+\mu)N - N^2} \tilde{E}_N((s,\infty);a,\mu),
\end{multline}
although the RHS is convergent for only a subset of the parameter values
for which the LHS converges (explicitly the RHS requires
Re$(a'+b') > -1$ and Re$(-N-b') > -1$ to be properly defined, while the LHS
only requires Re$(a'+b') > -1$). Note however that by replacing the
integrals over $\lambda_l \in [0,1]$ in (\ref{2.c})
by the Barnes double loop integral
\cite{WW65} the quantity $\tilde{E}_N((s,\infty);a,\mu)$ has meaning
for general values of $a$ and $\mu$. In the case $a'=0$,
$b'=k \in \zz_{>0}$, the LHS of (\ref{k2}) reads
\begin{equation}\label{k3}
  \Big \langle \prod_{l=1}^N  ( 1 + e^{-2 \pi i x_l} )^k
   e^{-s e^{2 \pi i x_l}}  \Big \rangle_{\rm CUE} =
  \Big \langle \det (1 + \bar{U})^k e^{-s {\rm Tr}(U)} 
  \Big \rangle_{U \in \rm CUE}.
\end{equation}
Like (\ref{21}) this latter average 
with the dimension $N$ replaced by $l$
is the generating function for certain
combinatorial objects \cite{TW99} (random words from an alphabet of $k$
letters with maximum increasing subsequence length constrained to be
less than or equal to $l$). 

As we have already stated, our interest is in the evaluation of
(\ref{2.c}) and (\ref{2.d}) as $\tau$-functions occurring in
Okamoto's theory of \PV. In some cases such evaluations, or closely
related evaluations, are already available in the literature.
One such case is (\ref{2.c}) and (\ref{2.d}) with $\mu=0$, for which
Tracy and Widom \cite{TW94c} 
(for subsequent derivations see \cite{ASV95, HS99, BD00}) have shown
\begin{align}
  \tilde{E}_N((0,s);a,0) & = \exp \int_0^s {\sigma (t) \over t} \, dt
  \label{2.e} \\
  \tilde{E}_N((s,\infty);a,0) & = \exp \Big ( - 
  \int_s^\infty {\sigma (t) \over t} \, dt \Big )
  \label{2.f}
\end{align}
where $\sigma(t)$ satisfies the Jimbo-Miwa-Okamoto $\sigma$-form of
the Painlev\'e V equation,
\begin{multline}\label{2.g}
  (t \sigma'')^2 - \left[ \sigma - t \sigma' + 2(\sigma')^2 +
  (\nu_0 + \nu_1 + \nu_2 + \nu_3) \sigma' \right]^2 \\
  + 4 (\nu_0 + \sigma')(\nu_1 + \sigma')(\nu_2 + \sigma')
  (\nu_3 + \sigma') = 0,
\end{multline}
with
\begin{equation}\label{30'}
  \nu_0 = 0, \quad \nu_1=0, \quad \nu_2 = N+a, \quad \nu_3 = N
\end{equation}
(or any permutation thereof, since (\ref{2.g}) is symmetrical in
$\{\nu_k\}$),
and subject to appropriate boundary conditions in each case.
For general $\mu$ the second integral in (\ref{2.c}) has been written
in an analogous form to the right hand side of (\ref{2.e}) by Adler
and van Moerbeke \cite{AV99b}, with the function corresponding to
$\sigma$ presented as the solution of a certain third order
differential equation. It is remarked in  \cite{AV99b} that this
equation can be reduced to a second order second degree equation,
known to be equivalent to (\ref{2.g}), using results due to
Cosgrove \cite{CS93,Co00} (see also the discussion in the final
section of \cite{WFC00}). The average (\ref{k3}) can be computed in
terms of Painlev\'e transcendents by making use of an identity
\cite{Fo93c} implying
\begin{equation}\label{32a}
 \Big\langle \det(1 + \bar{U})^k e^{-s {\rm Tr}(U)} \Big\rangle_{U \in {\rm CUE}}
  = e^{ks} \tilde{E}_k((0,s);N,0),
\end{equation}
(note the dual role played by $k$ and $N$ on the different sides of
(\ref{32a}))
then appealing to (\ref{2.e}) (direct evaluations have been given
by Tracy and Widom \cite{TW99} and Adler and van Moerbeke \cite{AV99b}).

The Okamoto $\tau$-function theory of \PV provides a unification and
extension of these results. 
Consider first (\ref{2.d}). Proposition \ref{p4} below gives 
\begin{align}\label{32}
  \tilde{E}_N((0,s);a,\mu) = {I_N(a\+ \mu) \over I_N(a)} 
  & \exp \int_0^s {V_N(t;a,\mu)\+ \mu N \over t} \, dt \nonumber
  \\
  = {I_N(\mu) \over I_N(a)} s^{Na} e^{-Ns}
  & \exp \Big( -\! \int_s^\infty {V_N(t;a,\mu)\+ Nt\m N(a\m \mu) \over t} \, dt
         \Big)
\end{align}
where $V_N(t;a,\mu)$ satisfies the Jimbo-Miwa-Okamoto $\sigma$ form of
\PV (\ref{2.g}) with
\begin{equation}\label{2.h3}
  \nu_0 = 0, \quad \nu_1 = - \mu, \quad
  \nu_2 = N + a, \quad \nu_3 = N,
\end{equation}
and subject to the boundary condition
\begin{equation}
  V_N(t;a,\mu) \mathop{\sim}\limits_{t \to \infty} -Nt + N(a - \mu)  -
  {N(N+\mu)a \over t} + O({1 \over t^2}).
\end{equation}
Regarding (\ref{2.c}), in Proposition \ref{pU} below we show
\begin{align}\label{35}
  \tilde{E}_N((s,\infty);a,\mu) 
  & = {J_N(a,\mu) \over I_N(a)} s^{(a+\mu)N + N^2}
  \exp \int_0^s {U_N(t;a,\mu) - aN - N^2 \over t} \, dt
  \nonumber \\
  & = s^{N\mu} 
    \exp \Big( - \int_s^\infty {U_N(t;a,\mu)  \over t} \, dt \Big)
\end{align}
where like $V_N(t;a,\mu)$, $U_N(t;a,\mu)$ satisfies (\ref{2.g}) with
parameters (\ref{2.h3}), but with the boundary condition
\begin{equation}
  U_N(t;a,\mu) \mathop{\sim}\limits_{t \to 0}
  aN + N^2 - N {a + N \over a + \mu + 2N}t.
\end{equation}
As well as $V_N(t;a,\mu)$ and $U_N(t;a,\mu)$ satisfying the differential
equation (\ref{2.g}), we show in Proposition \ref{LUE_diff} that they also 
satisfy third order difference equations in both the $a$ and $\mu$ variables.

By substituting the first equation of (\ref{35}) in (\ref{16}) and
(\ref{k2}) we obtain the evaluations
\begin{equation}
  \Big \langle e^{-s \sum_{j=1}^N \lambda_j} \Big \rangle_{\rm JUE}
  \Big |_{b \mapsto \mu}  =
  \exp \int_0^s {U_N(t;a,\mu) - aN - N^2 \over t} \, dt ,
  \label{37}
\end{equation}
\begin{multline}
  \Big \langle \prod_{l=1}^N e^{\pi i x_l(a'-b')}
       |1 + e^{2 \pi i x_l}|^{a'+b'} e^{-s e^{2 \pi i x_l}}
  \Big \rangle_{\rm CUE} \\
   = {M_N(a',b') \over M_N(0,0)}
  \exp \int_0^s {U_N(t;a,\mu) - aN - N^2 \over t} \, dt
  \Big |_{a = - (N+b') \atop \mu = a' + b'}. \label{38}
\end{multline}
Comparison of (\ref{37}) with (\ref{n1}) gives that for $G=O_N^{\pm},
USp(N)$,
\begin{equation}
  e^{\chi_{N^*} s} 
  \Big \langle e^{-s {\rm Tr}(U)} \Big \rangle_{U \in G}
  = e^{2sN^*} 
  \exp \int_0^{4s} {U_{N^*}(t;a,b) - aN^* - {N^*}^2 \over t} \, dt 
\end{equation}
with $(N^*,a,b)$ specified by (\ref{17'}) or (\ref{J4}) as appropriate.
In particular, with $G=USp(N)$ and thus $(N^*,a,b)$ given by (\ref{J4}),
substitution into (\ref{21}) shows
\begin{equation}\label{ply}
  P_l(s) = e^{-s^2/2} e^{2sl} \exp
  \int_0^{4s} {U_l(t;1/2,1/2) - l/2 - l^2 \over t} \, dt
\end{equation}

After recalling the statement below (\ref{R1}) and inserting the
proportionality constants in (\ref{u1u}) and  (\ref{u2}),
the evaluations (\ref{32}) and (\ref{35}) give
\begin{align}
  p_{\rm min}(s;a) \Big |_{N \mapsto N+1} 
  & = (N\+1) {I_N(a\+2) \over I_{N+1}(a)} s^a e^{-s} \exp \int_0^s
  {V_N(t;a,2)\+ 2N \over t} \,dt \label{1.45'} \\
  p_{\rm max}(s;a) \Big |_{N \mapsto N+1} 
  & = (N\+1) {I_N(a) \over I_{N+1}(a)} s^{a+2N} e^{-s}
  \exp \Big(\! -\!\int_s^\infty {U_N(t;a,2) \over t} \, dt \Big).
\end{align}
Also, recalling (\ref{13}) and the sentence below that equation, we see
from (\ref{32}) that 
\begin{multline}
  \rho(s) \Big |_{N \mapsto N + 1} \\
  = (N\+1) {I_N(a\+2) \over I_{N+1}(a)} s^{a}
  e^{-(N+1)s} \exp \Big( - \int_{-s}^0 {V_N(t;2;a) + aN \over t} \, dt
  \Big),
\end{multline}
where $\rho(s)$ denotes the eigenvalue density in the LUE.

According to Proposition \ref{p6.1}
\begin{equation}
  \tilde{E}_N((0,s);a,0;\xi) = \exp \int_0^s {W_N(t;a,0) \over t} \, dt
\end{equation}
where $W_N(t;a,0)$ satisfies (\ref{2.g}) with parameters (\ref{2.h3}),
$\mu$ set equal to zero, and boundary condition
\begin{equation}
  W_N(t;a,0)  \mathop{\sim}\limits_{t \to 0}
  \sim - \xi {\Gamma(N+a+1) \over \Gamma(N) \Gamma(a+1)\Gamma(a+2)} t^{a+1}.
\end{equation}
This result is known from \cite{TW94c}.

A generalization of the integral identity (\ref{32a}) is derived in the
context of the Painlev\'e theory in Proposition \ref{p7} below. We
remark that this generalization is in fact a special case of a still
more general identity, known from an earlier study \cite{Fo93c}, involving
an arbitrary parameter $\beta$. For $\beta = 1$ an average in the Laguerre
orthogonal ensemble is related to an average in the circular symplectic
ensemble, while for $\beta = 4$ an average in the Laguerre symplectic
ensemble is related to an average in the circular orthogonal ensemble.

The Laguerre ensemble permits four scaled, large $N$ limits. These are
\cite{Fo93a,BFP98,Jo99a,Jo01}
\begin{align}
  s \mapsto s/4N, & \; N \to \infty \label{1a}\\
  s \mapsto 4N + 2(2N)^{1/3} s, & \; N \to \infty \label{1b}\\
  a = (\gamma - 1)N, \: \: s \mapsto N(1- \sqrt{\gamma})^2
   - \nu_-(N) s, & \; N \to \infty,  \label{1c}\\
  a = (\gamma - 1)N, \: \: s \mapsto N(1+ \sqrt{\gamma})^2
   + \nu_+(N) s, & \; N \to \infty,  \label{1c'}
\end{align}
where in (\ref{1c}), (\ref{1c'}) we require $\gamma > 1$ and
\begin{equation}\label{1d}
   \nu_-(N) = \Big[N(\sqrt{\gamma} - 1)(1 - {1 \over \sqrt{\gamma}})
  \Big]^{\frac{1}{3}}, \quad
   \nu_+(N) = \Big[N(\sqrt{\gamma} + 1)(1 + {1 \over \sqrt{\gamma}})
  \Big]^{\frac{1}{3}}.
\end{equation}
The first gives the limiting distributions at the hard edge, so called because
the eigenvalue density is strictly zero on one side. The second, third and
fourth
give the limiting distribution at the soft edge, so called because there is
a non-zero density for all values of the new coordinates, but with a fast
decrease on one side.

The soft edge limit for the quantities (\ref{1.a}) and (\ref{1.b}) in the
GUE was
studied in our work \cite{FW00}, where evaluations in terms of
solutions of the general Jimbo-Miwa-Okamoto $\sigma$-form of \PII
\begin{equation}\label{46'}
  (u'')^2 + 4u' \Big ( (u')^2 - tu' + u \Big ) - \alpha^2 = 0.
\end{equation} 
were given. From either of the limiting procedures (\ref{1b}), 
(\ref{1c}) or (\ref{1c'}) we reclaim the results of  \cite{FW00}
(because no new functional forms appear we will not discuss this case
further in subsequent sections).
In particular, it follows from the second equation in (\ref{35}) that 
\begin{align}\label{rec}
  \tilde{E}^{\rm soft}(s;\mu) 
  & = \lim_{s \mapsto 4N + 2(2N)^{1/3}s \atop N \to \infty}
  C  e^{-\mu s/2} \tilde{E}_N((s,\infty);a,\mu)
  \\
  & = \tilde{E}^{\rm soft}(s_0;\mu) \exp\int_{s_0}^s u(t;\mu)\, dt,
  \nonumber
\end{align}
where 
\begin{equation}\label{ut}
  u(t,\mu) = \lim_{N \to \infty}
  {2 (2N)^{1/3} \over 4N}
  \Big ( U_N(t;a,\mu) + N \mu - {\mu \over 2} t \Big )
  \Big |_{t \mapsto 4N + 2(2N)^{1/3} t}.
\end{equation}
Making the replacements
\begin{gather}
  \sigma \mapsto \sigma - N \mu + {\mu \over 2}t, \nonumber\\
  t \mapsto 4N + 2(2N)^{1/3}t, \quad
  \sigma(4N+2(2N)^{1/3}t) \mapsto (2N)^{2/3}u(t;\mu) \nonumber
\end{gather}
in (\ref{2.g}) with parameters (\ref{2.h3}), and equating terms of
order $N^2$ in the equation (which is the leading order) shows
$u(t;\mu)$ satisfies (\ref{46'}) with $\alpha = 
\mu $. We know from \cite{FW00} that (\ref{46'}) is to be
solved subject to the boundary condition 
\begin{equation}\label{64}
  u(t;\mu) \mathop{\sim}\limits_{t \to -\infty} {1 \over 4} t^2
  + {4 \mu^2 - 1 \over 8t} + {(4\mu^2 - 1)(4 \mu^2 - 9) \over 64 t^4} +
\cdots
\end{equation}

As an application, (\ref{ut}) can be used to compute the scaled limit
\begin{equation*}
  P(s) := \lim_{l \to \infty} P_l\Big (l - {1 \over 2} (2l)^{1/3}s \Big ).
\end{equation*}
Imposing the condition that $P(s) \to 1$ as $s \to \infty$ (which fixes
the constants), a short calculation using (\ref{ut}) in
(\ref{ply}) shows
\begin{equation}\label{pf}
  P(s) = \exp \Big ( - \int_s^\infty( u(-t;1/2) - t^2/4) \, dt \Big ).
\end{equation}
This is interesting because it has been proved by Baik and Rains
\cite{BR99b} that
\begin{equation}\label{pf1}
  P(2^{1/3}s) = F_1(s),
\end{equation}
where $F_1(s)$ is the cumulative distribution function for the largest
eigenvalue in the scaled, infinite Gaussian orthogonal ensemble.
Tracy and Widom \cite{TW96} (see \cite{Fo99b} for a simplified derivation)
have shown that
\begin{equation}\label{tw8}
  F_1(s)  = e^{- {1 \over 2} \int_s^\infty (t-s) q^2(t) \, dt}
e^{{1 \over 2} \int_s^\infty q(t) \, dt}
\end{equation}
where with Ai$(t)$ denoting the Airy function,
$q(t)$ is the solution of the non-linear equation
\begin{equation}\label{3}
  q'' = 2q^3 + tq,
\end{equation}
(Painlev\'e II equation with $\alpha = 0$)
subject to the boundary condition
\begin{equation}\label{4}
  q(t) \sim - {\rm Ai}(t) \qquad {\rm as} \qquad t \to \infty.
\end{equation}
By equating (\ref{pf}) and (\ref{pf1}) we obtain as an alternative
to (\ref{tw8}) the formula
\begin{equation}\label{tw}
  F_1(2^{-1/3}s) =
  \exp \Big ( - \int_s^\infty( u(-t;1/2) - t^2/4) \, dt \Big ).
\end{equation}
However the boundary condition (\ref{64}) gives
$u(-t;1/2) - t^2/4 \sim 0$ which is vacuous, whereas the boundary
condition according to the requirement (\ref{4}) is
\begin{equation*}
  2^{-1/3} {d \over dt} ( u(-t;1/2) - t^2/4)
  \Big |_{t \mapsto 2^{-1/3}t}\,
  \mathop{\sim}\limits_{t \to \infty}
   {1 \over 2} \Big[  - {\rm Ai}'(t) +
  \Big ( {\rm Ai}(t) \Big )^2 \Big].
\end{equation*}
The significance of (\ref{tw}) from the viewpoint
of gap probabilities in random matrix ensembles as
$\tau$-functions for Hamiltonians associated with Painlev\'e
systems will be discussed in a separate publication \cite{FW01}.

For the hard edge
scaling (\ref{1a}), define the scaled version of (\ref{2.d}) by
\begin{equation}\label{1.54}
  \tilde{E}^{\rm hard}(s;a,\mu) :=
  \lim_{s \mapsto s/4N \atop N \to \infty} \Big (
  {I_N(a) \over I_N(a\+ \mu)} \tilde{E}_N((0,s);a,\mu) \Big ).
\end{equation}
In the case $\mu=0$, it is known from the work of Tracy and Widom
\cite{TW94b} that
\begin{equation}\label{1.55}
  \tilde{E}^{\rm hard}(s;a,0) = \exp \Big ( - \int_0^s {\sigma_B(t) \over
  t} \, dt \Big )
\end{equation}
where $\sigma_B(t)$ satisfies the Jimbo-Miwa-Okamoto $\sigma$-form of the 
Painlev\'e III equation
\begin{equation}\label{sa}
  (t \sigma'')^2 - v_1 v_2 (\sigma')^2 + \sigma'(4 \sigma' - 1)
  (\sigma - t \sigma') - {1 \over 4^3} (v_1 - v_2)^2 = 0
\end{equation}
with
\begin{equation}\label{sa1}
  v_1 = a, \qquad v_2 = a.
\end{equation}
It follows from Proposition \ref{p12} below that for general $\mu$
\begin{equation}\label{sa2}
  \tilde{E}^{\rm hard}(s;a,\mu) =
   \exp 
  \Big ( - \int_{0}^s {\sigma(t) + \mu(\mu + a)/2 \over t} \, dt \Big )
\end{equation}
where $\sigma$ satisfies (\ref{sa}) with
\begin{equation}\label{sa3}
  v_1 = a+ \mu, \qquad v_2 = a - \mu
\end{equation}
and subject to the boundary condition
\begin{equation}\label{sa4}
  \sigma(t) \mathop{\sim}\limits_{t \to \infty}
  {t \over 4} - { at^{1/2} \over 2} +
  \Big ( {a^2 \over 4} - {\mu^2 \over 2} \Big ). 
\end{equation}
In Proposition \ref{HE_diff} we show that $ v(t;a,\mu) =-\sigma(t)-\mu(\mu+a)/2 $
satisfies third order difference equations in both $a$ and $\mu$.

One application of (\ref{sa2}) is to the evaluation of
\begin{equation}\label{rt}
  p_{\rm min}^{\rm hard}(s;a) := \lim_{N \to \infty} {1 \over 4N}
  p_{\rm min} \Big ( {s \over 4N}; a \Big ) .
\end{equation}
First we note that the explicit evaluation of (\ref{2.h1}) is
$\prod_{j=1}^N \Gamma(1+j) \Gamma(a+j)$, which together with the asymptotic
formula $\Gamma(x+a)/ \Gamma(x) \sim x^a$ for $x \to \infty$ shows that
\begin{equation*}
  \lim_{N \to \infty} {1 \over 4N} (N\+ 1)
  {I_N(a\+ 2) \over I_{N+1}(a)} s^a e^{-s} \Big |_{s \mapsto s/4N}
  = {s^a \over 2^{2a+2} \Gamma(a+1) \Gamma(a+2)}.
\end{equation*}
Using this formula and (\ref{sa2}) we can take the large $N$ limit
in (\ref{1.45'}) as required by (\ref{rt}) to conclude
\begin{equation}\label{rt1}
  p_{\rm min}^{\rm hard}(s;a) =
  {s^a \over 2^{2a+2} \Gamma(a+1) \Gamma(a+2)}
  \exp \Big ( - \int_0^s {\sigma(t)+a+2 \over t} \, dt \Big )
\end{equation}
where $\sigma(t)$ satisfies (\ref{sa}) with
\begin{equation*}
  v_1 = a+2, \qquad v_2 = a-2
\end{equation*}
and is subject to the boundary condition (\ref{sa4}) with $\mu=2$. Similarly,
it follows that the hard edge density is given by
\begin{equation}\label{1.73a}
  \rho^{\rm hard}(s) = {s^a e^{-s/4} \over 2^{2a+2} \Gamma(a+1) \Gamma(a+2)}
  \exp  \Big (  \int_{-s}^0 {\sigma(t) + a(a+2)/2 \over t} \, dt \Big )
\end{equation}
where $\sigma(t)$ satisfies (\ref{sa}) with
\begin{equation*}
  v_1 = 2+a, \qquad v_2=2-a
\end{equation*}
and is subject to the boundary condition (\ref{sa4}) with
$(a,\mu) = (2,a)$.

The result (\ref{sa2}) also has relevance to the CUE. 
This occurs via the scaled limit of (\ref{32a}), which has been
shown \cite{Fo93c} to imply
for $a \in \zz_{\ge 0}$, $\mu \in \zz$, the identity
\begin{equation}
  \tilde{E}^{\rm hard}(t;a,\mu) \propto
  e^{-t/4} t^{-\mu a/2} \Big \langle
  e^{{1 \over 2} \sqrt{t} {\rm Tr} (U + \bar{U})}
  (\det U)^{-\mu} \Big \rangle_{U \in {\rm CUE}_a}
\end{equation}
(eq.~(\ref{4.31}) below; the proportionality constant is specified in
(\ref{4.32})). It thus follows that this CUE${}_a$ average can be evaluated
in terms of the transcendent $\sigma(t)$ in (\ref{sa2}) (in the case
$\mu = 0$ this has been shown directly in
the previous works \cite{
TW99a,AV99b}).

This concludes the summary of our results. In the next section (Section 2)
an overview of the Okamoto $\tau$-function theory of the Painlev\'e V
equation is given. In particular a sequence of $\tau$-functions
$\tau[n]$, each of which can be characterised as the solution of a certain
second order second degree equation, is shown to satisfy a Toda lattice
equation. The fact that a special choice of initial parameters in the
sequence permits the evaluations $\tau[0]=1$ and $\tau[1]$ an explicit
function of $t$ (a confluent hypergeometric function) then implies 
that the general member of the sequence $\tau[n]$ is given 
explicitly as a Wronskian type determinant depending on $\tau[1]$. In
Section 3 the  Wronskian type determinants from Section 2 are evaluated
in terms of the multiple integrals (\ref{2.d}) and (\ref{2.c}), thus
identifying them as $\tau$-functions and implying their characterization
as the solution of a non-linear second order second degree equation.

In Section 4 the Okamoto $\tau$-function theory of the Painlev\'e III
equation is revised, and again a $\tau$-function sequence is identified
which can be shown to coincide with a quantity in the LUE
(explicitly $\tilde{E}^{\rm hard}(s;a,\mu)$ for general $\mu$ and
$a \in \zz_{\ge 0}$). By scaling results from Section 3 it is shown
that for general $\mu$ and general $a$,  $\tilde{E}^{\rm hard}(s;a,\mu)$
as specified in (\ref{1.54}) is a $\tau$-function in the \PIII theory.
Concluding remarks relating to the boundary conditions
(\ref{sa4}) and (\ref{64}) are given in Section 5.
 
\section{Overview of the Okamoto $\tau$-function theory of \PV}
\setcounter{equation}{0}
\subsection{The Jimbo-Miwa-Okamoto $\sigma$-form of \PV}
As formulated in \cite{Ok87}, the Okamoto $\tau$-function theory of the
fifth Painlev\'e equation \PV is based on the Hamiltonian system
$ \{Q,P,t,K\} $
\begin{multline}\label{6.1}
  tK = Q(Q-1)^2P^2 
       - \Big[ (v_2 - v_1)(Q-1)^2 - 2(v_1+v_2)Q(Q-1) +tQ \Big]P \\
       + (v_3 - v_1)(v_4 - v_1)(Q-1)\qquad
\end{multline}
where the parameters $v_1,\dots,v_4$ are constrained by
\begin{equation}\label{6.1a}
  v_1 + v_2 + v_3 + v_4 = 0.
\end{equation}
The relationship of (\ref{6.1}) to \PV can be seen by eliminating $P$
in the Hamilton equations
\begin{equation}\label{6.1b}
   Q' = {\partial K \over \partial P}, \qquad
   P' = - {\partial K \over \partial Q}.
\end{equation}
One finds that $Q$ satisfies the equation
\begin{equation}\label{6.1c}
   y'' = \Big( {1 \over 2y} + {1 \over y - 1} \Big) (y')^2
         - {1 \over t} y' 
         + {(y-1)^2 \over t^2} \Big( \alpha y + {\beta \over y} \Big) 
         + \gamma{y \over t} + \delta{y(y+1) \over y-1}
\end{equation}
with
\begin{equation}
  \alpha = {1 \over 2}(v_3 - v_4)^2, \quad
  \beta = - {1 \over 2} (v_2 - v_1)^2, \quad
  \gamma = 2 v_1 + 2v_2 - 1, \quad
  \delta = - {1 \over 2}.
\end{equation}
This is the general \PV equation with $\delta = - {1 \over 2}$ (recall that
the general \PV equation with $\delta \ne 0$ can be reduced to the case
with $\delta = - {1 \over 2}$ by the mapping $t \mapsto \sqrt{-2 \delta} t$).
Note that the first of the Hamilton equations (\ref{6.1b}) implies
\begin{equation}
  tQ' = 2Q(Q-1)^2P - \Big\{ (v_2 - v_1)(Q-1)^2 - 2(v_1+v_2)Q(Q-1) +tQ \Big\},
\end{equation}
which is linear in $P$,
so using this equation to eliminate $P$ in (\ref{6.1}) shows that
$tK$ can be expressed as an explicit rational function of $Q$ and $Q'$. 

Of fundamental importance in the random matrix application is the fact that
$tK$ (or more conveniently a variant of $tK$ obtained by adding a term linear
in $t$) satisfies a second order second degree equation.

\begin{prop}\label{p1}
\cite{JM81,Ok87} With $K$ specified by (\ref{6.1}), define the auxiliary
Hamiltonian $h$ by
\begin{equation}\label{6.4}
   h = tK + (v_3 - v_1)(v_4 - v_1) - v_1t - 2v_1^2.
\end{equation}
The auxiliary Hamiltonian $h$ satisfies the differential equation
\begin{equation}\label{6.4a}
   (th'')^2 - \Big (h - th' + 2(h')^2\Big )^2 + 4 \prod_{k=1}^4(h'+v_k) = 0.
\end{equation}
\end{prop}

\begin{proof}
Following \cite{Ok87}, we note from (\ref{6.1}),
(\ref{6.1a}) and the Hamilton equations (\ref{6.1b}) that  
\begin{align}\label{6.4'}
   h'   & = -QP - v_1 \nonumber \\
   th'' & = Q(Q^2-1)P^2 - \Big[ (v_3+v_4-2v_1)Q^2 - v_2 + v_1 \Big]P
        \\
        & \qquad + (v_3 - v_1)(v_4 - v_1)Q. \nonumber
\end{align}
We see from these equations that
\begin{equation}\label{6.5}
   th'' = Q(h'+v_3)(h'+v_4) + P(h'+v_2),
\end{equation}
while the first of 
these equations together with (\ref{6.1}) and (\ref{6.4}) shows
\begin{equation}\label{6.6}
   h - th' + 2(h')^2 = Q(h'+v_3)(h'+v_4) - P(h'+v_2).
\end{equation}
Squaring both sides of (\ref{6.5}) and (\ref{6.6}) and subtracting, and
making further use of the first equation in (\ref{6.4'}) gives (\ref{6.4a}).
\end{proof}

\medskip
Actually the differential equation obtained by Jimbo and Miwa
\cite{JM81} is not precisely (\ref{6.4a}), but rather a variant obtained
by writing
\begin{equation*}
  h = \sigma^{(j)} - v_jt - 2 v_j^2, \quad j=1, \ldots ,4\ ,
\end{equation*}
or equivalently
\begin{equation}\label{6.5a}
  \sigma^{(j)} = tK + (v_3-v_1)(v_4-v_1) - (v_1 - v_j)t - 2(v_1^2 - v_j^2).
\end{equation}
With this substitution (\ref{6.4a}) coincides with
the Jimbo-Miwa-Okamoto $\sigma$ form of \PV (\ref{2.g}) with 
\begin{equation}\label{6.5c}
  \{ \nu_0,\nu_1,\nu_2,\nu_3\} = 
  \{ v_1 - v_j, v_2 - v_j, v_3-v_j, v_4-v_j \} 
\end{equation}
(because (\ref{2.g}) is symmetrical in $\{\nu_k\}$ the ordering in the
correspondence (\ref{6.5c}) is arbitrary).

The $\tau$-function is defined in terms of the Hamiltonian by
\begin{equation}\label{2.11'}
  K =: {d \over dt} \log \tau.
\end{equation}
It then follows from (\ref{6.5a}) that
\begin{equation}\label{6.5d}
  \sigma^{(j)} = t {d \over dt} \log \Big(
   e^{-(v_1 - v_j)t} t^{(v_3 - v_1)(v_4 - v_1) - 2(v_1^2 - v_j^2)} \tau \Big).
\end{equation}
Now, according to the Okamoto theory the $\tau$-function corresponding to
some particular sequences of parameter values can be calculated recursively
in an explicit determinant form. The above theory tells us that these
$\tau$-functions can also be characterised as the solutions of a
non-linear differential equation. The determinant solutions are a consequence
of a special invariance property of the \PV system which we will now
summarise.

\subsection{B\"acklund transformations}
A B\"acklund transformation of (\ref{6.1}) is a mapping, in fact a birational
canonical transformation 
\begin{equation*}
  T(\mathbf{v};q,p,t,H) = (\bar{\mathbf{v}};\bar{q},\bar{p},\bar{t},\bar{H}),
\end{equation*}
where $\bar{\mathbf{v}}, \bar{q}, \bar{p}, \bar{t}, \bar{H}$ are functions of
$\mathbf{v},q,p,t,H$, such that the \PV coupled system (\ref{6.1b}) is satisfied
in the variables $(\bar{\mathbf{v}}; \bar{q}, \bar{p}, \bar{t})$. Formally
\begin{equation*}
  dp \wedge dq - dH \wedge dt = 
  d\bar{p} \wedge d\bar{q} - d\bar{H} \wedge d\bar{t}.
\end{equation*}
In general,
the significance of such a transformation is that it allows an infinite
family of solutions of the \PV system to be obtained from one seed solution.
From our perspective a key feature is the property
\begin{equation}\label{u1}
  T^{-1}(H) = H \Big |_{\mathbf{v} \mapsto T\cdot\mathbf{v}}
\end{equation}
which holds for some particular operators $T$ which have a shift action
on the parameter vector $ \mathbf{v}=(v_1,v_2,v_3,v_4) $ whose components
are referred to the standard basis.  Explicitly,
following \cite{Ok87}, we are interested in the transformation $T_0$ which
has the property (\ref{u1}) with the action on the parameters
\begin{equation}\label{u1'}
  T_0\cdot\mathbf{v} = 
  \Big( v_1-{1\over 4}, v_2-{1\over 4}, v_3-{1\over 4}, v_4+{3\over 4} \Big).
\end{equation}
Although the existence of $T_0$ was established in \cite{Ok87} (termed the 
parallel transformation $ l(\mathbf{v}) $), as was the
fact that $T_0(t) = t$, explicit formulas for $T_0(P)$ and $T_0(Q)$ 
were not presented.

It is now realised \cite{Wa98,K99} that the required formulas are more
readily forthcoming by developing the \PV theory starting from the
symplectic coordinates and Hamiltonian $ \{q,p,t,H\} $ specified by
\begin{equation}\label{u2'}
  tH := q(q-1)p(p+t) - (v_2-v_1+v_3-v_4)qp + (v_2-v_1)p + (v_1-v_3)tq.
\end{equation}
The coordinates and Hamiltonians constituting the two charts (\ref{6.1}) and 
(\ref{u2'}) are related by the coordinate transformation \cite{Wa98}
\begin{align}\label{u3}
  (q-1)(Q-1) & = 1 \nonumber \\
  (q-1)p + (Q-1)P & = v_3-v_1 \nonumber \\
  tK & = tH + (v_3-v_1)(v_2-v_4)
\end{align}
so in particular $1-1/q$ satisfies the Painlev\'e V equation (\ref{6.1c}).

The Hamiltonian (\ref{u2'}) is associated with a set of four coupled 
first order equations in symmetric variables \cite{Ad94,NY98} which 
admit an extended type $A^{(1)}_3$ affine Weyl group 
$ \tilde{W}_a = W_a(A^{(1)}_3)\rtimes Z_4 $ as B\"acklund transformations. 
The generators of the group 
$\tilde{W}_a = \langle s_0, s_1, s_2, s_3, \pi \rangle$ obey the 
algebraic relations
\begin{equation}
\begin{split}
  & s_i^2 = 1, \quad s_i s_{i\pm 1} s_i = s_{i\pm 1} s_i s_{i\pm 1}, 
  \quad s_is_j = s_js_i \quad (j\neq i,i\pm 1),
  \\
  & \pi s_i = s_{i+1} \pi, \quad \pi^4 = 1,
  \quad (i,j=0,\dots,3, \: s_4:= s_0).
\end{split}
\label{awgA3}
\end{equation}
The parameters of \PV can also be characterised by the level sets of the 
root vectors $ \alpha_0, \alpha_1, \alpha_2, \alpha_3 $ (with
$ \alpha_0 + \alpha_1 + \alpha_2 + \alpha_3 = \delta $)
of the $A^{(1)}_3$ root system, and are dual to the vector parameter 
$ \mathbf{v} $ in the following way
\begin{equation*}
  \alpha_1(\mathbf{v}) := v_2-v_1, \quad 
  \alpha_2(\mathbf{v}) := v_1-v_3, \quad
  \alpha_3(\mathbf{v}) := v_3-v_4 \ . 
\end{equation*}
In general by the convention of \cite{NY99} the action of a group operation 
$ T $ on the roots is given in terms
of its action on the components of $ \mathbf{v} $ by
\begin{equation}
  T(\alpha)\cdot\mathbf{v} = \alpha(T^{-1}\cdot\mathbf{v}) \ .
\end{equation}
Following \cite{K99} the action of the  B\"acklund transformations on the root
system $ \alpha $ and coordinates $q$, $p$ is given in Table \ref{t1}. 
Note that with $T_0$ defined by (\ref{u1'}),
\begin{equation*}
  T_0(\alpha) = (\alpha_0 -1, \alpha_1, \alpha_2, \alpha_3 +1),
\end{equation*}
so according to Table \ref{t1} we  have
\begin{equation}\label{dd}
  T_0 = s_3 s_2 s_1\pi, \quad T^{-1}_0 = \pi^{-1} s_1 s_2 s_3.
\end{equation}

As is the case for the dynamical system associated with \PIV there is a 
symmetric form for the system describing the \PV transcendent \cite{NY98}, 
and is expressed in terms of the four variables
\begin{equation}
\begin{split}
   f_0 & = {p+t \over \sqrt{t}} \\
   f_1 & = \sqrt{t}q \\
   f_2 & = -{p \over \sqrt{t}} \\
   f_3 & = \sqrt{t}(1-q),
\end{split}
\label{A3_fvars}
\end{equation}
with the constraints $ f_0+f_2 = f_1+f_3 = \sqrt{t} $. Using these variables the
B\"acklund transformations take the simple and generic forms of 
\begin{equation}
\begin{split}
  s_i(\alpha_j) & = \alpha_j-a_{ij}\alpha_i, \quad
  \pi(\alpha_j) = \alpha_{j+1}, \\
  s_i(f_j)      & = f_j+u_{ij}{\alpha_i \over f_i},  \quad
  \pi(f_j) = f_{j+1}, \\
\end{split}
\label{A3_BT}
\end{equation}
where the Cartan matrix $ A = (a_{ij}) $ and orientation matrix $ U = (u_{ij}) $
are defined by 
\begin{equation}
  A = \left[ \begin{array}{rrrr} 2&-1&0&-1\\-1&2&-1&0\\0&-1&2&-1\\-1&0&-1&2
             \end{array} \right] ,\quad
  U = \left[ \begin{array}{rrrr} 0&1&0&-1\\-1&0&1&0\\0&-1&0&1\\1&0&-1&0
             \end{array} \right] .
\end{equation}

Let us now compute the action of $ T^{-1}_0 $ on the Hamiltonian $ tH $.
First we can easily verify from Table \ref{t1} that
\begin{align}\label{2.18p}
  s_0(tH) & = tH + \alpha_0{t \over p+t} + \alpha_0(\alpha_2 -1) \nonumber \\
  s_1(tH) & = tH + \alpha_1 t + \alpha_1 \alpha_3 \nonumber \\
  s_2(tH) & = tH - \alpha_2 t + \alpha_2(\alpha_0 -1) \nonumber \\
  s_3(tH) & = tH + \alpha_1 \alpha_3 \nonumber \\
  \pi(tH) & = tH + (q-1)p - \alpha_2 t. 
\end{align}
It follows from these formulas, Table \ref{t1} and (\ref{dd}) that
\begin{equation}\label{TH}
  T^{-1}_0(tH) = tH \Big|_{\mathbf{v} \mapsto T_0\cdot\mathbf{v}} =
  tH \Big|_{\mathbf{\alpha} \mapsto T^{-1}_0\cdot\mathbf{\alpha}} = tH + qp.
\end{equation}
According to the relations in (\ref{u2'}) between $tH$ and $tK$,
$q, p$ and $Q, P$, we therefore have
\begin{equation}\label{2.17'}
\begin{split}
  T^{-1}_0(tK) & = tK - Q(Q-1)P + (v_3-v_1)(Q - 1)
  \\
  & = tK \Big |_{\mathbf{v} \mapsto T_0\cdot\mathbf{v}} =
  tK \Big |_{\mathbf{\alpha} \mapsto T^{-1}_0\cdot\mathbf{\alpha}},
\end{split}
\end{equation}
which is the result deduced indirectly in \cite{Ok87}.
Note from (\ref{u3}) that 
\begin{align}\label{2.20'}
  T^{-1}_0(Q) & = 1 + {1 \over T^{-1}_0(q) - 1} \nonumber \\
  T^{-1}_0(P) & = { (v_3-v_1) - (T^{-1}_0(q) - 1) T^{-1}_0(p) 
                    \over T^{-1}_0(q) - 1 }
\end{align}
where $T^{-1}_0(q)$ and $T^{-1}_0(p)$ can be deduced from 
Table \ref{t1} using (\ref{dd}).

\begin{table}
\begin{center}
\begin{tabular}{|c||c|c|c|c|c|c|}\hline
 & $\alpha_0$ & $\alpha_1$ & $\alpha_2$ & $\alpha_3$ & $p$ & $q$
 \\ \hline
 $s_0$ & $- \alpha_0$ & $\alpha_1 + \alpha_0$ & $\alpha_2$ &
 $\alpha_3 + \alpha_0$ &
 $p$ & $q + {\displaystyle\alpha_0 \over \displaystyle p+t}$
 \\
 $s_1$ & $\alpha_0 + \alpha_1$ & $- \alpha_1$ & $\alpha_2 + \alpha_1$ &
 $\alpha_3$ & 
 $p - {\displaystyle\alpha_1 \over \displaystyle q}$ & $q$
 \\
 $s_2$  & $\alpha_0$ &  $\alpha_1 + \alpha_2$ & $- \alpha_2$ &
 $\alpha_3 + \alpha_2$  & 
 $p$ & $q + {\displaystyle\alpha_2 \over \displaystyle p}$
 \\
 $s_3$ & $\alpha_0 + \alpha_3$ & $\alpha_1$ &  $\alpha_2 + \alpha_3$ &
 $-  \alpha_3$ & 
 $p - {\displaystyle\alpha_3 \over \displaystyle q-1}$ & $q$
 \\
 $\pi$ & $\alpha_1$ & $\alpha_2$ & $\alpha_3$ &  $\alpha_0$ & 
 $t(q-1)$ & $- {\displaystyle p \over \displaystyle t}$
 \\ \hline
\end{tabular}
\end{center}
\caption{\label{t1}  B\"acklund transformations relevant for the \PV 
Hamiltonian (\ref{u2'}).}
\end{table}

\subsection{Toda lattice equation}
Introduce the sequence of Hamiltonians
\begin{equation}\label{tns}
  K[n] := K \Big|_{\mathbf{v} 
                   \mapsto (v_1 -n/4, v_2 -n/4, v_3 -n/4, v_4 +3n/4)}
\end{equation}
and let $\tau[n]$ denote the corresponding $\tau$-functions so that
\begin{equation}\label{tn}
  K[n] = {d \over dt} \log \tau[n].
\end{equation}
We have already remarked that a crucial feature of $K[n]$ from the
viewpoint of application to random matrix theory is the fact that it
(or more precisely the quantity (\ref{6.5a})) satisfies the differential
equation (\ref{2.g}). An equally crucial feature for application to
random matrix theory is the recurrence satisfied by $\{\tau[n]\}$.

\begin{prop} \cite{Ok87,K99} 
The $\tau$-function sequence $\tau[n]$,
corresponding to the parameter sequence $(v_1-n/4,v_2-n/4,v_3-n/4, v_4+3n/4)$, 
obeys the Toda lattice equation
\begin{equation}\label{toda1}
  \delta^2 \log \bar{\tau}[n] =
  {\bar{\tau}[n-1] \bar{\tau}[n+1] \over \bar{\tau}^2[n]}, 
  \qquad \delta := t{d \over dt}
\end{equation}
where
\begin{equation}\label{toda2}
\bar{\tau}[n] := t^{n^2/2} e^{(v_4 - v_1 + n)t} \tau[n].
\end{equation}
\end{prop}

\begin{proof}
Following \cite{K99} we consider the sequence of Hamiltonians
$H[n]$ associated with the given parameter sequence and the
Hamiltonian (\ref{u2'}). Now, with $\hat{\tau}[n]$ denoting the
corresponding $\tau$-function, it follows from (\ref{u3}) that
\begin{equation*}
  \tau[n] = t^{(v_3-v_1)(v_2-v_4-n)} \hat{\tau}[n].
\end{equation*}
Since (\ref{toda1}) is unchanged by 
the replacement $\bar{\tau}[n] \mapsto t^{a+bn} \bar{\tau}[n]$, it suffices
to show that (\ref{toda1}) is satisfied with $\hat{\tau}[n]$ replacing
$ \tau[n]$ in (\ref{toda2}).

From the definitions
\begin{equation}\label{lhs}
  \delta \log {\hat{\tau}[n-1] \hat{\tau}[n+1] \over \hat{\tau}^2[n]}
  = \Big( T^{-1}_0(tH[n]) - tH[n] \Big) -
    \Big( tH[n] - T_0(tH[n]) \Big).
\end{equation}
On the other hand, it follows from (\ref{TH}) that
\begin{multline}\label{A}
  \Big( T^{-1}_0(tH[n]) - tH[n] \Big) -
  \Big( tH[n] - T_0(tH[n]) \Big) \\
   = q[n]p[n] - T_0(q[n])T_0(p[n]).
\end{multline}
With $T_0^{-1}$ specified from (\ref{dd}), explicit formulas for
$T_0(q[n])$ and $T_0(p[n])$ can be deduced from
Table \ref{t1}, and it can be verified using the Hamilton equations
for $H$ that the RHS of (\ref{A}) is equal to
\begin{multline*}
  \delta \log \Big( q[n](q[n]-1)p[n] + (v_1-v_3)q[n] + (v_4-v_1+n) \Big) 
  \\
  = \delta \log{d \over dt} \Big( tH[n] + (v_4-v_1+n)t \Big) =
    \delta \log{d \over dt} \delta \log\Big(
        e^{(v_4-v_1+n)t} \hat{\tau}[n] \Big).
\end{multline*}
Equating this with the LHS of (\ref{lhs}) gives a formula equivalent to
(\ref{toda1}).
\end{proof}

\medskip
\subsection{Classical solutions}
The Toda lattice equation (\ref{toda1}) is a second order recurrence, and
so requires the values of $\bar{\tau}[0]$ and $\bar{\tau}[1]$ for the
sequence members $\bar{\tau}[n]$, $(n \ge 2)$ to be specified.
It was shown by Okamoto \cite{Ok87} that for special choices of the
parameters, corresponding to the chamber walls in the underlying $A_3$
root lattice, the \PV system admits a solution with $\tau[0]=1$ and
$\tau[1]$ equal to a confluent hypergeometric function.

\begin{prop} \cite{Ok87} For the special choice of parameters
\begin{equation}\label{v14}
   v_1 = v_4
\end{equation}
it is possible to choose $\tau[0]=1$. Furthermore, the first member
$\tau[1]$ of the $\tau$-function sequence (\ref{tn}) then
satisfies the confluent hypergeometric equation
\begin{equation}\label{v15}
   t(\tau[1])'' + (v_3-v_2+1+t)(\tau[1])' + (v_3-v_1)\tau[1] = 0.
\end{equation}
\end{prop}

\begin{proof}
Write $ Q=Q[0], P=P[0] $. In the case $v_1 = v_4$ we see from 
(\ref{6.1}) that it is possible to choose 
\begin{equation}\label{v15'}
  P=0, \quad K=0,
\end{equation}
the latter allowing us to take $\tau[0]=1$. With (\ref{v15'}) the equations
(\ref{2.17'}), (\ref{6.1}) and (\ref{tn}) then give
\begin{equation}\label{v16}
  t{d \over dt} \log\tau[1](t) = (v_3-v_1)(Q-1).
\end{equation}
The quantity $Q$ must satisfy the Hamilton equation
\begin{equation}\label{v17}
  tQ' = {\partial tK \over \partial P} \Big|_{v_1 = v_4 \atop P=0} =
        -(v_2-v_1)(Q-1)^2 - (v_3-v_2)Q(Q-1) - tQ
\end{equation}
where use has been made of (\ref{6.1a}). Substituting (\ref{v16}) in
(\ref{v17}) gives (\ref{v15}).
\end{proof}

\medskip
The two linearly independent solutions of (\ref{v15}) are
\begin{align}\label{v18}
  \tau^{\rm a}[1](t) 
  & = {}_1 F_1(v_3-v_1, v_3-v_2+1; -t) \nonumber \\
  & = {\Gamma(v_3-v_2+1) \over \Gamma(v_3-v_1)\Gamma(v_1-v_2+1)}
       \int_0^1 e^{-tu} u^{v_3-v_1-1}(1-u)^{v_1-v_2} \, du
\end{align}
and
\begin{align}\label{v19}
  \tau^{\rm s}[1](t) 
  & = e^{-t} \psi(1+v_1-v_2, v_3-v_2+1; t) \nonumber \\
  & = {e^{-t} \over \Gamma(1+v_1-v_2)}
       \int_0^\infty e^{-tu} u^{v_1-v_2}(1+u)^{v_3-v_1-1} \, du
\end{align}
(the superscripts ``a" and ``s" denote analytic and singular respectively,
and refer to the neighbourhood of $t=0$).

As noted by Okamoto \cite{Ok87}, it is a classical result 
that the solution of the Toda
lattice equation in the case $\bar{\tau}[0]=1$ is given in terms of 
$\bar{\tau}[1](t)$ via the determinant formula
\begin{equation}\label{det}
  \bar{\tau}[n] = \det [ \delta^{j+k} \bar{\tau}[1](t) ]_{j,k=0,\dots,n-1}.
\end{equation}
The formulas (\ref{v18}) and (\ref{v19}) substituted in (\ref{toda2})
with $v_4 = v_1$ and thus reading
\begin{equation}\label{78'}
  \bar{\tau}[n] = t^{n^2/2} e^{nt} \tau[n]
\end{equation}
give the explicit form of $ \bar{\tau}[1](t)$, and thus we have two distinct
explicit determinant formulas for $\bar{\tau}[n]$. 

\medskip
\subsection{Schlesinger Transformations}
We are now in a position to derive difference equations for the dynamical
quantities of the \PV system which are consequences of the Schlesinger 
transformations or shift operators corresponding to translations by the 
fundamental weights of the $ A^{(1)}_3 $ lattice
\begin{equation}
  T_0 = s_3s_2s_1\pi, \quad T_1 = \pi s_3s_2s_1, \quad
  T_2 = s_1\pi s_3s_2, \quad T_3 = s_2s_1\pi s_3. 
\end{equation}
It is well known that these difference equations can be identified as discrete 
Painlev\'e equations \cite{ROG-2000} satisfying integrable criteria analogous 
to the continuous ones. Here we briefly demonstrate this and find a difference 
equation for the Hamiltonians. 

\begin{prop}\cite{ROG-2000}
The Schlesinger transformation of the \PV system for the shift operator 
$ T^{-1}_0 $ generating the parameter sequence \\
$ (\alpha_0+n, \alpha_1, \alpha_2, \alpha_3-n) $
with $ n \in \mathbb{Z} $,
corresponds to the second order difference equation of the discrete Painlev\'e 
type, d\PIV, namely
\begin{equation}
\begin{split}
  x_{n}+x_{n-1} & = {t \over y_{n}} + {\alpha_3-n \over 1-y_{n}} \\
   y_{n}y_{n+1} & =  t{ x_{n}+\frac{1}{2}\alpha_1+\alpha_0+n
                        \over x^2_{n}-\frac{1}{4}\alpha^2_1 }
  \qquad n \geq 1 \ ,
\end{split}
\label{dPIV-T0}
\end{equation}
where $ x_{n} = f_0[n]f_1[n]-\frac{1}{2}\alpha_1 $ and 
$ y_{n} = \sqrt{t}/f_1[n] $.
With $ qp[n] = tH[n\+ 1]-tH[n] $, the Hamiltonian times $ t $, $ tH[n] $,
satisfies the third order difference equation
\begin{equation}
\begin{split}
 - tqp[n] = 
   & \Big\{
     (\alpha_0\+ n\+ qp[n])[tH[n]+(\alpha_3\m n)qp[n]-\alpha_2 t]
   \\
   & \quad 
     + (\alpha_2\+ qp[n\+ 1])
       [tH[n]+(\alpha_3\m n)qp[n]+(\alpha_0\+ \alpha_1\+ n)t] \Big\}
   \\
   \div & \Big\{
      tH[n]+(\alpha_3\m n)qp[n]
       +(\alpha_3\m n\m 1)(\alpha_2\+ qp[n\+ 1])-\alpha_2 t \Big\}
   \\
   \times & \Big\{
     (qp[n]\m \alpha_3\+ n)[tH[n]+(1\m \alpha_0\m n)qp[n]+\alpha_1 t]
   \\
   & \quad
     + (qp[n\m 1]\m \alpha_1)
       [tH[n]+(1\m \alpha_0\m n)qp[n]-(\alpha_2\+ \alpha_3\m n)t] \Big\}
   \\
   \div & \Big\{
      tH[n]+(1\m \alpha_0\m n)qp[n]
       +(1\m \alpha_0\m n)(qp[n\m 1]\m \alpha_1)+\alpha_1 t \Big\}.
\end{split}
\label{dPIV-Ham}
\end{equation}
\end{prop}

\begin{proof}
The Schlesinger transformations for the shift operator $ T_0 $
that are relevant for the identification are
\begin{align}
  T^{-1}_0(f_0)
  & = f_3 - \cfrac{\alpha_0}{f_0}
          + \cfrac{\alpha_0 \+\alpha_1 \+\alpha_2}
                  {f_2 + \cfrac{\alpha_0 \+\alpha_1}{f_1 
                       + \cfrac{\alpha_0}{f_0}}}
  \ ,\label{Sxfm-T0:a} \\
  T^{-1}_0(f_1)
  & = f_0 - \cfrac{\alpha_0 \+\alpha_1}{f_1 + \cfrac{\alpha_0}{f_0}}
  \ ,\label{Sxfm-T0:b} \\
  T^{-1}_0(f_3)
  & = f_2 + \cfrac{\alpha_0 \+\alpha_1}{f_1 + \cfrac{\alpha_0}{f_0}}
  \ .\label{Sxfm-T0:c} 
\end{align}
Using (\ref{Sxfm-T0:c}) in (\ref{Sxfm-T0:a}) along with the constraint to eliminate
$ f_3 $ we can rewrite the latter equation as
\begin{equation}
   f_1[n] + \cfrac{\alpha_0 \+ n}{f_0[n]} =
   \sqrt{t} - f_0[n\+ 1] + \cfrac{n\+ 1\m \alpha_3}{\sqrt{t}-f_1[n\+ 1]}.
\end{equation}
This identity can then be employed in (\ref{Sxfm-T0:b}) to arrive at
\begin{equation}
   f_0[n]f_1[n] +  f_0[n\m 1]f_1[n\m 1] = \alpha_1 + \sqrt{t}f_1[n]
   + {n\m \alpha_3 \over \sqrt{t}-f_1[n]}f_1[n] ,
\end{equation}
which is the first of the coupled equations (\ref{dPIV-T0}).
The second equation of the set (\ref{dPIV-T0}) follows immediately from
(\ref{Sxfm-T0:b}) rewritten to read
\begin{equation}
   {1\over f_1[n]f_1[n\+ 1]} = {1\over f_0[n]f_1[n]}
   {f_0[n]f_1[n] \+ \alpha_0 \+ n \over f_0[n]f_1[n] \m \alpha_1} .
\end{equation}
The simplest way to derive the difference equation for $ tH[n] $ is to recast 
the Schlesinger transformations in terms of canonical variables $ q, p $,
\begin{align}
  tq[n\+ 1]
  & = t+p[n] - \cfrac{\alpha_0 \+ \alpha_1 \+ n}{
                      q[n] + \cfrac{\alpha_0 +n}{t+p[n]}}
  \ ,\label{Sqp-T0:a} \\
  p[n\+ 1]
  & = -tq[n] - \cfrac{(\alpha_0 +n)t}{t+p[n]}
             + \cfrac{(n\+ 1\m \alpha_3)t}{-p[n] 
             + \cfrac{\alpha_0 \+ \alpha_1 \+ n}{
                      q[n] +\cfrac{\alpha_0 \+ n}{t+p[n]}}}
  \ ,\label{Sqp-T0:b} \\
  tq[n\m 1]
  & = -p[n] + \cfrac{n\m \alpha_3}{1-q[n]}
            + \cfrac{1\m \alpha_0 \m n}{q[n] 
            + \cfrac{\alpha_2 \+ \alpha_3 \m n}{
                     p[n] +\cfrac{\alpha_3 -n}{1-q[n]}}}
  \ ,\label{Sqp-T0:c} \\
   t+p[n\m 1]
  & = tq[n] + \cfrac{(\alpha_2 \+ \alpha_3 \m n)t}{
                      p[n] +\cfrac{\alpha_3 \m n}{1-q[n]}}
  \ .\label{Sqp-T0:d}
\end{align}
Considering the first two equations (\ref{Sqp-T0:a},\ref{Sqp-T0:b}) we find that
\begin{equation}
   qp[n\+ 1] +\alpha_2 = 
   {[\alpha_0 \+ n\+ q[n](t\+ p[n])][qp[n](t\+ p[n])- \alpha_1 p[n]+\alpha_2 t]
   \over 
    p[n][\alpha_1 \m q[n](t\+ p[n])] + (\alpha_0 \+ \alpha_1 \+ n)t}
\end{equation}
and expressing this in terms of the Hamiltonian yields
\begin{equation}
   qp[n\+ 1]\+ \alpha_2 = 
  -{[tq[n]\+ \alpha_0\+ n\+ qp[n]][tH[n]\+(\alpha_3 \m n)qp[n]-\alpha_2 t]
   \over 
     tH[n]\+ (\alpha_3 \m n)qp[n]
          \+ (\alpha_3 \m n\m 1)tq[n]\+(\alpha_0 \+\alpha_1 \+ n)t} .
\end{equation}
In an analogous way we find for the down-shifted product
\begin{equation}
   qp[n\m 1]\m \alpha_1 =
   {[p[n]\+ \alpha_3 \m n\m qp[n]][tH[n]+(1\m \alpha_0 \m n)qp[n]+\alpha_1 t]
   \over 
     tH[n]\+ (1\m \alpha_0 \m n)qp[n]
          \m (1\m \alpha_0 \m n)p[n]\m (\alpha_2 \+ \alpha_3 \m n)t} .
\end{equation}
Now the up-shifted equation can be solved for $ q $ and the down-shifted one
for $ p $ allowing their product $ qp[n] $ to be expressed in terms of just
the Hamiltonian and other members of the $ qp $ sequence. The final result is
(\ref{dPIV-Ham}).
\end{proof}

\section{Application to the finite LUE}
\setcounter{equation}{0}
\subsection{The case $N=1$}
Comparing the case $N=1$ of the integrals (\ref{2.d}) and (\ref{2.c}) with
those in (\ref{v18}) and (\ref{v19}) we see that
\begin{align*}
  \tilde{E}_1((0,s);a,\mu) & = C s^{a+\mu+1} \tau^{\rm s}[1](s)
  \Big |_{v_1 - v_2 = \mu \atop v_3 - v_1 = a+1} \\
  \tilde{E}_1((s,\infty);a,\mu) & = C s^{a+\mu+1} \tau^{\rm a}[1](s)
  \Big |_{v_1 - v_2 = \mu \atop v_3 - v_1 = a+1} 
\end{align*}
Recalling (\ref{6.5d}), we thus have that
\begin{align*}
  t {d \over dt} \log \Big( t^{-\mu} \tilde{E}_1((0,t);a,\mu) \Big) 
  & = \sigma^{{\rm s}(1)}(t)\\
  t {d \over dt} \log \Big(
  t^{-\mu} \tilde{E}_1((t,\infty);a,\mu) \Big)
  & = \sigma^{{\rm a}(1)}(t),
\end{align*}
where both $\sigma^{{\rm s}(1)}(t)$ and $\sigma^{{\rm a}(1)}(t)$
satisfy the Jimbo-Miwa-Okamoto $\sigma$ form of \PV (\ref{2.g}) with
\begin{equation*}
  \nu_0 = 0, \quad \nu_1 = - \mu, \quad \nu_2 = a+1, \quad \nu_3=1.
\end{equation*}
Note that in the case $\mu=0$ this is consistent with (\ref{30'}).

\subsection{The general $N$ case}
Although it is not at all immediately obvious, the $n \times n$
determinant formed by substituting (\ref{v18}) and (\ref{v19}) in
(\ref{det}) can be identified with the general $n$ cases of the integrals
(\ref{2.d}) and (\ref{2.c}) respectively. Consider first
(\ref{det}) with initial value (\ref{v18}).

\begin{prop}\label{pU}
Let $\bar{\tau}[n]$ be specified by the determinant formula (\ref{det})
with 
\begin{equation}\label{f1}
\bar{\tau}[1](t) = t^{1/2} e^t {}_1 F_1(v_3\m v_1,v_3\m v_2\+ 1;-t).
\end{equation}
Then we have
\begin{equation}\label{89'}
  \bar{\tau}[n] \propto 
     t^{n^2/2} \prod^{n}_{l=1} \int_0^{1} du_{l} \,
     e^{tu_{l}} u_{l}^{v_1-v_2} (1-u_{l})^{v_3-v_1-n}
    \prod_{1\leq j<k \leq n}(u_k - u_j)^2.
\end{equation}
It follows from this that
\begin{equation}
  \tilde{E}_N((t,\infty);a,\mu) = C t^{(a+\mu)N+N^2}\tau[N](t)
\end{equation}
and
\begin{equation}\label{pv1}
  t {d \over dt} \log \Big( t^{-N \mu} \tilde{E}_N((t,\infty);a,\mu) \Big)
  = U_N(t;a,\mu) 
\end{equation}
where $U_N(t;a,\mu)$ satisfies the Jimbo-Miwa-Okamoto $\sigma$ form of
the \PV \\ differential equation (\ref{2.g}) with
\begin{equation}\label{pv}
  \nu_0 = 0, \quad \nu_1=-\mu, \quad \nu_2 = N+a, \quad \nu_3 = N,
\end{equation}
subject to the boundary condition
\begin{equation}\label{91'}
  U_N(t;a,\mu) \mathop{\sim}\limits_{t \to 0}
    aN + N^2 - N {a + N \over a + \mu + 2N}t.
\end{equation}
Equivalently $U_N(t;a,\mu)$ is equal to the auxiliary Hamiltonian
(\ref{6.5a}) with $j=1$ and parameters (\ref{pv}).
\end{prop}

\begin{proof}
First observe that if $\{\bar{\tau}[n]\}$ satisfies the
Toda equation (\ref{toda1}) with $\bar{\tau}[0] = 1$, then
$\{t^{nc} \bar{\tau}[n]\}$ is also a solution which is given by
the determinant formula (\ref{det}) with $\bar{\tau}[1](t)
\mapsto t^{c} \bar{\tau}[1](t) $. Choosing $c = -1/2$ and
substituting (\ref{f1}) for $\bar{\tau}[1](t)$ we thus have
\begin{equation}\label{b1}
  t^{-n/2}  \bar{\tau}[n] = \det
  \Big [ \delta^{j+k} {}_1 F_1(v_1\m v_2\+ 1, v_3\m v_2\+ 1;t)
  \Big ]_{j,k=0,\dots,n-1}
\end{equation}
where use has been made of the Kummer relation
\begin{equation*}
  {}_1 F_1(a,c;-t) = e^{-t} {}_1 F_1(c-a,c;t).
\end{equation*}
Our first task is to use elementary row and column operations to
eliminate the operator $\delta^{j+k}$ in (\ref{b1}). Starting with
elementary column operations, in column $k$ ($k = n-1,n-2,\dots,1$ in
this order) make use of the identity
\begin{equation}\label{12.1}
  \delta {}_1 F_1(a,c;t) = 
  a \Big( {}_1 F_1(a+1,c;t) - {}_1 F_1(a,c;t) \Big)
\end{equation}
and add $(v_1-v_2+1)$ times column $k-1$. This gives
\begin{multline*}
  t^{-n/2} \bar{\tau}[n] \propto
  \det \Big[ \delta^j {}_1 F_1(v_1\m v_2\+ 1, v_3\m v_2\+ 1;t) \;\ldots \\
             \delta^{j+k-1}  {}_1 F_1(v_1\m v_2\+ 2, v_3\m v_2\+ 1;t) 
             \;\ldots
       \Big]_{j=0,\dots,n-1 \atop k=1,\dots,n-1}.
\end{multline*}
Next, in column $k$ ($k = n-1,n-2, \dots,2$ in this order) make use of the
identity (\ref{12.1}) again and add $(v_1-v_2+2)$ times column $k-1$
to obtain
\begin{align}\label{96'}
  t^{-n/2} \bar{\tau}[n] \propto \det \Big[
  & \delta^j {}_1F_1(v_1\m v_2\+ 1,v_3\m v_2\+ 1;t) \\
  & \delta^j {}_1F_1(v_1\m v_2\+ 2,v_3\m v_2\+ 1;t) \ldots \nonumber \\
  & \delta^{j+k-2} {}_1F_1(v_1\m v_2\+ 3,v_3\m v_2\+ 1;t) \quad\ldots \nonumber 
    \Big]_{j=0,\dots,n-1 \atop k=2,\dots,n-1}.
\end{align}
Further use of (\ref{12.1}) in an analogous fashion gives
\begin{equation}\label{13.1}
  t^{-n/2} \bar{\tau}[n] \propto \det \Big [
  \delta^j {}_1F_1(v_1\m v_2\+ 1\+ k, v_3\m v_2\+ 1;t) \Big ]_{j,k=0,\dots,n-1}.
\end{equation}

At this stage we use elementary row operations to eliminate the
operation $\delta^j$ in (\ref{13.1}). In row $j$ ($j=n-1,n-2,\dots,1$ in
this order) make use of the identity
\begin{equation}\label{12.2}
  {d \over dt} {}_1 F_1(a,c;t) = {a-c \over c}
  {}_1 F_1(a,c+1;t) +  {}_1 F_1(a,c;t)
\end{equation}
and subtract row $j-1$ to get
\begin{multline*}
   t^{-n/2} \bar{\tau}[n] \propto \\
  \det \left[
  \begin{array}{c}
    {}_1 F_1(v_1\m v_2\+ 1\+ k, v_3\m v_2\+ 1;t) \\
    (v_3 - v_1-k) \delta^{j-1} t
    {}_1 F_1(v_1\m v_2\+ 1\+ k, v_3\m v_2\+ 2;t)
  \end{array}
  \right]_{j=1,\dots,n-1 \atop k=0,\dots,n-1}.
\end{multline*}
Next, in row $j$ ($j=n-1,n-2,\dots,2$ in this order) make use of the
identity (\ref{12.2}) again and subtract 2 times row $j-1$ to obtain
\begin{multline*}
   t^{-n/2} \bar{\tau}[n] \propto \\
   t\det \left[
  \begin{array}{c}
    {}_1 F_1(v_1\m v_2\+ 1\+ k, v_3\m v_2\+ 1;t) \\
    (v_3 - v_1-k) t
    {}_1 F_1(v_1\m v_2\+ 1\+ k, v_3\m v_2\+ 2;t) \\
    (v_3 - v_1-k)_2 \delta^{j-2} t^2
    {}_1 F_1(v_1\m v_2\+ 1\+ k, v_3\m v_2\+ 3;t)
  \end{array}
  \right]_{j=2,\dots,n-1 \atop k=0,\dots,n-1}.
\end{multline*}
Further use of (\ref{12.2}) in an analogous fashion gives
\begin{multline*}
  t^{-n/2}  \bar{\tau}[n] \propto \\
  t^{n(n-1)/2} \det \Big[ (v_3\m v_1\m k)_j
  {}_1 F_1(v_1\m v_2\+ 1\+ k, v_3\m v_2\+ 1\+ j; t) \Big]_{j,k=0,\dots,n-1}
\end{multline*}
thus eliminating entirely the operation $\delta$.

We now substitute the integral representation for 
${}_1 F_1$ deducible from (\ref{v18}) to obtain
\begin{align}\label{14.1}
  t^{-n/2} \bar{\tau}[n] 
  & \propto t^{n(n-1)/2}
    \det \Big[ \!\int_0^1 e^{tu} u^{k+v_1-v_2} (1\m u)^{v_3-v_1-1+j-k}
      du \Big]_{j,k=0,\dots,n-1} 
  \nonumber\\
  & = t^{n(n-1)/2} \prod^{n}_{l=1} \int_0^{1} du_{l} \,
      e^{tu_{l}} u_{l}^{v_1-v_2} (1\m u_{l})^{v_3-v_1-n}
  \nonumber\\
  & \qquad\times \det \Big[ u_{j+1}^k(1\m u_{j+1})^{n+j-k-1}
                      \Big]_{j,k=0,\dots,n-1}.
\end{align} 
Regarding the determinant in the last line of this expression, we have
\begin{multline*}
  \det \Big[ u_{j+1}^k(1 - u_{j+1})^{n+j-k-1} \Big]_{j,k=0,\dots,n-1} \\
  = \prod_{j=0}^{n-1}(1-u_{j+1})^j \det [ u_{j+1}^k]_{j,k=0,\dots,n-1},
\end{multline*}
where the equality follows by adding the $k\!+\!1$-th column to the
$k$-th $(k=n-1,\dots,l)$ in a sequence of sweeps $l=1,\ldots,n-1$.
All factors other than the determinant in the multidimensional integral
are symmetric in $\{u_j\}$ so we can symmetrise this term without
changing the value of the integral (apart from a constant factor
$n!$ which is not relevant to the present discussion). Noting that
\begin{equation}
{\rm Sym} \Big ( \prod_{j=0}^{n-1}(1-u_{j+1})^j
                 \det[ u_{j+1}^k]_{j,k=0,\dots,n-1}
          \Big ) = {\pm \over n!} \prod_{j < k} (u_j - u_k)^2,
\end{equation}
for some sign $\pm$, and substituting in (\ref{14.1}) gives (\ref{89'}). 
\begin{align*}
  t^{-n/2} \bar{\tau}[n] 
  \propto t^{n(n-1)/2} \prod^{n}_{l=1}
  \int_0^{1} du_{l} \, e^{tu_{l}} u_{l}^{v_1-v_2} (1-u_{l})^{v_3-v_1-n}
  \prod_{j < k} (u_j - u_k)^2.
\end{align*}
Recalling (\ref{78'}), changing variables
$u_j \mapsto 1 - u_j$ and comparing with the final integral in
(\ref{2.c}) we thus have that
\begin{equation*}
  \tau[N](t) \propto t^{-(a+\mu)N - N^2}
  \tilde{E}_N((t,\infty);a,\mu)
\end{equation*}
provided the parameters are given by (\ref{pv}). The result
(\ref{pv1}) now follows by substituting this result in (\ref{6.5d})
with $j=1$. For the boundary condition (\ref{91'}), we see
from (\ref{pv1}) and (\ref{2.c}) that
\begin{equation*}
  U_N(t;a,\mu) \mathop{\sim}\limits_{t \to 0}
  aN + N^2 - t {J_N(a,\mu)[\sum_{j=1}^N \lambda_j] \over J_N(a,\mu)},
\end{equation*}
where $J_N(a,\mu)[\sum_{j=1}^N \lambda_j]$ denotes the integral
(\ref{2.h1}) with an additional factor of $\sum_{j=1}^N \lambda_j$
in the integrand. Noting that $\sum_{j=1}^N \lambda_j$ can be
written as a ratio of alternants allows the integral to computed
using the method of orthogonal polynomials and leads to the result
(\ref{91'}).
\end{proof}

\medskip
Next we consider (\ref{det}) with initial value (\ref{v19}).

\begin{prop}\label{p4}
Let $\bar{\tau}[n]$ be specified by the determinant formula (\ref{det})
with
\begin{equation}\label{f11}
  \bar{\tau}[1](t) = t^{1/2} \psi(v_1\m v_2\+ 1,v_3\m v_2\+ 1;t).
\end{equation}
Then we have
\begin{equation}\label{c1}
  \bar{\tau}[n] \propto 
   t^{n^2/2} \prod^{n}_{l=1} \int_0^\infty du_{l} \, 
   e^{-tu_{l}} u_{l}^{v_1-v_2} (1+ u_{l})^{v_3-v_1-n}
    \prod_{1\leq j<k \leq n}(u_k - u_j)^2.
\end{equation}
It follows from this that
\begin{equation}
  \tilde{E}_N((0,t);a,\mu) = C t^{(a+\mu)N+N^2}\tau[N](t)
\end{equation}
and
\begin{equation}\label{c2}
  t {d \over dt} \log \Big ( t^{-N \mu} \tilde{E}_N((0,t);a,\mu) \Big )
  = V_N(t;a,\mu)
\end{equation}
where $V_N(t;a,\mu)$ is equal to the auxiliary
Hamiltonian (\ref{6.5a}) with $j=1$ and parameters (\ref{pv}), and so
satisfies the Jimbo-Miwa-Okamoto form of
\PV (\ref{2.g}) with parameters (\ref{pv}). The latter is to be solved  
subject to the boundary condition
\begin{equation}\label{c3}
  V_N(t;a,\mu) \mathop{\sim}\limits_{t \to \infty}
  -Nt + N(a-\mu) - {N(N+\mu)a \over t} + O(1/t^2).
\end{equation}
\end{prop}

\begin{proof}
Analogous to (\ref{b1}) we have
\begin{equation}\label{92a}
  t^{-n/2}  \bar{\tau}[n] = \det
  \Big [ \delta^{j+k} \psi (v_1\m v_2\+ 1, v_3\m v_2\+ 1;t)
  \Big ]_{j,k=0,\dots,n-1}
\end{equation}
We now adopt the identical strategy as used in the proof of Proposition
\ref{pU}, with the identities (\ref{12.1}) and (\ref{12.2}) replaced by
\begin{equation*}
  \delta \psi(a,c;t) = 
  a \Big( (a-c+1) \psi (a+1,c;t) - \psi (a,c;t) \Big)
\end{equation*}
and
\begin{equation*}
  {d \over dt} \psi (a,c;t) = 
  \psi (a,c;t) - \psi (a,c+1;t)
\end{equation*}
respectively. The results (\ref{c1}) and (\ref{c2}) are then obtained by
repeating the working which led to (\ref{89'}) and
(\ref{pv1}).

For the boundary condition, we see from (\ref{c2}) and (\ref{2.d}) that
\begin{equation*}
  V_N(t;a,\mu) \mathop{\sim}\limits_{t \to \infty}
  -Nt + N(a-\mu) - {a I_N(\mu)[\sum_{j=1}^N \lambda_j] \over tI_N(\mu)},
\end{equation*}
where $I_N(\mu)[\sum_{j=1}^N \lambda_j]$ denotes the first integral 
in (\ref{2.h1})
with an extra factor of $\sum_{j=1}^N \lambda_j$ in the integrand. 
This integral can be computed by changing variables $\lambda_j \mapsto
\epsilon \lambda_j$ in the definition of $I_N(\mu)$, differentiating
with respect to $\epsilon$, and setting $\epsilon = 1$.
\end{proof}

In relation to (\ref{1.14a}) we note from the fact that (\ref{v15})
is linear, with linearly independent solutions (\ref{v18}) and
(\ref{v19}), that
\begin{equation}\label{1.14c}
  \tau^{\xi}[1](t) = {e^{-t} \over \Gamma(1+v_1-v_2)}
  \Big ( \int_{-1}^\infty - \xi \int_{-1}^0 \Big )
  e^{-t u} u^{v_1 - v_2} (1 + u)^{v_3 - v_1 - 1} \, du
\end{equation}
is proportional to the most general solution of (\ref{v15}).
Substituting (\ref{1.14c}) in (\ref{toda2}) with $n=1$ and
$v_4 = v_1$, forming (\ref{det}), and simplifying by
noting that the considerations
of the proof of Proposition \ref{p4} (excluding the discussion of the
boundary condition) remain valid independent of the value of
$\xi$, we obtain the following result.

\begin{prop}\label{p6.1}
Let $\bar{\tau}[n]$ be specified by the determinant formula (\ref{det})
with $\bar{\tau}[1]$ proportional to $t^{1/2}$ times (\ref{1.14c}).
Then
\begin{equation}
  \tilde{E}_N((0,s);a,\mu;\xi) = C t^{(a+\mu) N + N^2}
\tau[N](t)
\end{equation}
and
\begin{equation}
  t {d \over dt} \Big ( t^{-N \mu} \tilde{E}_N((0,t);a,\mu;\xi) \Big )
= W_N(t;a,\mu)
\end{equation}
where $W_N(t;a,\mu)$, like $U_N(t;a,\mu)$ from Proposition \ref{pU}
and $V_N(t;a,\mu)$ from Proposition \ref{p4}, is equal to the auxiliary
Hamiltonian (\ref{6.5a}) with $j=1$ and parameters (\ref{pv}),
and so satisfies the Jimbo-Miwa-Okamoto form of PV (\ref{2.g})
with parameters (\ref{pv}). In the case $\mu = 0$, with $\rho(t)$ denoting
the eigenvalue density of the LUE, the latter is to be solved subject
to the boundary condition
\begin{equation}
  {W_N(t;a,0) \over t} \mathop{\sim}\limits_{t \to 0}
  - \xi \rho(t) \sim - \xi {\Gamma(N+a+1) \over \Gamma(N) \Gamma(a+1)
  \Gamma(a+2)} t^a.
\end{equation}
\end{prop} 

\subsection{A relationship between transcendents}
The evaluations (\ref{1.45'}) for $p_{\rm min}(s;a) \Big |_{N \mapsto
N + 1}$ and (\ref{32}) for $\tilde{E}_N((0,s);a,0)$ substituted into
(\ref{R0}) imply that
\begin{equation}\label{i1}
  V_N(t;a,2) = - (a + 1 + 2N) + t + V_{N+1}(t;a,0) + t
  {V_{N+1}'(t;a,0) \over V_{N+1}(t;a,0)}.
\end{equation} 
In our previous study \cite{FW00} we encountered an analogous identity
relating some \PIV transcendents. Like (\ref{i1}), the \PIV identity in
\cite{FW00} was discovered using a relation of the type (\ref{R0}). A
subsequent independent derivation using B\"acklund transformations was
also found and presented. Likewise the identity (\ref{i1}) can be derived
from the formulas (\ref{2.18p}) and Table \ref{t1}.
An identical formula relates $ U_N(t;a,2) $ and $ U_{N+1}(t;a,0) $ which
can be understood from the evaluation of 
$p_{\rm max}(s;a) \Big |_{N \mapsto N + 1}$ and $\tilde{E}_N((s,\infty);a,0)$.

\begin{prop}
Introduce the shift operators $T_2$ and $T_3$, which have the action on
the parameters $\mathbf{v}$ specified by
\begin{equation}\label{T1T3}
\begin{split}
  T_2\cdot\mathbf{v} 
  & = (v_1+{3 \over 4},v_2-{1 \over 4},v_3-{1 \over 4},v_4-{1 \over 4}),
  \\
  T_3\cdot\mathbf{v}
  &  = (v_1-{1 \over 4},v_2-{1 \over 4},v_3+{3 \over 4},v_4-{1 \over 4})
\end{split}
\end{equation}
and consequently from (\ref{2.17'}) and Table \ref{t1} the representation
\begin{equation}\label{T3b}
  T_2 = s_1 \pi s_3 s_2, \qquad T_3 = s_2 s_1 \pi s_3 .
\end{equation}
Then we have
\begin{equation}\label{T1a}
  T_3 T_2^{-1} V_{N+1}(t;a,0) = V_N(t;a,2)
\end{equation}
and use of (\ref{2.18p}) and  Table \ref{t1} on the LHS reduces this to
(\ref{i1}).
\end{prop}

\begin{proof}
We take advantage of the arbitrariness in (\ref{6.5c}) which
with $\{ \nu_i \}$ given by (\ref{30'}) with $N \mapsto N + 1$ allows us
to write
\begin{equation}\label{T1b}
  V_{N+1}(t;a,0) = 
  \sigma^{(1)} \bigg |_{{\scriptstyle v_2 - v_1 = a+N+1
                   \atop \scriptstyle v_3 - v_1 = 0}
                   \atop \scriptstyle v_4 - v_1 = N+1}.
\end{equation}
The actions (\ref{T1T3}) then imply
\begin{equation*}
  T_3 T_2^{-1} V_{N+1}(t;a,0) =  
  \sigma^{(1)} \bigg |_{{\scriptstyle v_2 - v_1 = a+N
                   \atop \scriptstyle v_3 - v_1 = -2}
                   \atop \scriptstyle v_4 - v_1 = N},
\end{equation*}
which recalling (\ref{2.h3}) and (\ref{6.5c}) is the equation (\ref{T1a}).

To evaluate the action of $T_3 T_2^{-1}$ on the LHS of (\ref{T1a}), first
note from (\ref{6.5a}) and the final equation in (\ref{u3}) that for
general parameters
\begin{equation*}
  \sigma^{(1)} = tH + (v_3-v_1)(v_2-v_1).
\end{equation*}
We remark that the simplifying feature of arranging the parameters so that
$ \alpha_2 = v_3-v_1 = 0$ in (\ref{T1b}) is that (\ref{u2'}) takes the reduced 
form
\begin{equation}\label{T1c}
  tH \Big|_{\alpha_2 = 0} = 
  p \Big[ q(q-1)(p+t) - (\alpha_1 +\alpha_3)q + \alpha_1 \Big].
\end{equation}
The explicit formula (\ref{T1c}) will become important later on. For now
see seek the action of $T_3 T_2^{-1}$ on $\sigma^{(1)}$ for general
values of the parameters. 
First we note that the generators of the
extended type $A^{(1)}_3$ affine Weyl group possess the algebraic properties
(\ref{awgA3}) which can be used in the formula for $T_3 T_2^{-1}$
implied by (\ref{T3b}) to show that
\begin{equation*}
  T_3 T_2^{-1} = s_2 s_3 s_1 s_0 s_1 s_3.
\end{equation*}
Now, it follows from (\ref{2.18p}) and Table \ref{t1} that 
\begin{equation}\label{T1d}
  s_0 s_1 s_3(tH) = tH + \alpha_0{t \over p+t}
                      + \alpha_0(\alpha_2 \m 1) + (\alpha_1\+ \alpha_0)t.
\end{equation}
Let us now consider separately the second term in (\ref{T1d}). Use of
Table \ref{t1} shows
\begin{equation*}
  s_3 s_1(\alpha_0{t \over p+t}) =
  {(1 \m \alpha_2)t q(q-1) \over q(q-1)(p+t) - \alpha_1(q-1) - \alpha_3 q}.
\end{equation*}
Furthermore, we see from Table \ref{t1} that in the special circumstance
$\alpha_2 = 0$ (i.e.~$v_3 - v_1 = 0$), $s_2$ acts like the identity on
$p$ and $q$ so we have
\begin{align}\label{tr1}
  s_2 s_3 s_1(\alpha_0{t \over p+t}) \Big|_{\alpha_2 = 0} 
  & = {tq(q-1) \over q(q-1)(p+t) - \alpha_1(q-1) - \alpha_3 q} \nonumber \\
  & = t{(tH)' \over tH} \Big|_{\alpha_2 = 0}
\end{align}
where the final equality follows from (\ref{T1c}) and (\ref{u2'}).
For the remaining terms in (\ref{T1d}), we see from (\ref{2.18p})
and Table \ref{t1} that
\begin{equation*}
  s_2 s_3 s_1 \left[ tH + \alpha_0( \alpha_2 \m 1) + (\alpha_1 \+ \alpha_0)t
              \right]\Big|_{\alpha_2 = 0}
  = tH + t - \alpha_0.
\end{equation*}
The above results together with the simple formula
\begin{equation*}
  T_3 T_2^{-1}(-\alpha_1\alpha_2) \Big|_{\alpha_2 = 0} =
  2(1\m \alpha_1) = 2(1-v_2+v_1)
\end{equation*}
imply
\begin{multline}\label{tr3}
  T_3 T_2^{-1} \sigma^{(1)} \Big|_{v_3 - v_1 = 0} \\
  = tH  \Big|_{v_3 - v_1 = 0} + t + 1 - (v_2-v_1) - (v_4-v_1) + 
    t{(tH)' \over tH} \Big|_{v_3 - v_1 = 0}.
\end{multline}
Substituting the values of $v_2-v_1$ and $v_4-v_1$ from (\ref{T1b}) gives
(\ref{i1}). 
\end{proof}

\subsection{Difference Equations}
We know the logarithmic derivatives of the $ \tau$-functions $ U_N(t;a,\mu) $
and $ V_N(t;a,\mu) $ satisfy the second order second degree differential 
equation (\ref{2.g}). Here we will utilise the Schlesinger transformation
theory to show that they also satisfy third order difference equations in both 
$ a $ and $ \mu $ variables.

\begin{prop}\label{LUE_diff}
The logarithmic derivative $ U_N(t;a,\mu) $ satisfies a third order difference
equation in the variable $ a $
\begin{equation}
\begin{split}
   -t(\uU-U) = 
  & \phantom{div} \Big\{
    (\uuU-U\+ a\+ \mu\+ 1)\big[ N(\mu\+ t)+(a\+ 1)U-a\uU \big] 
  \\
  & \qquad\qquad
    +(a\+ 1)t\big[ \uuU-\uU-N \big] \Big\}
  \\
  & \div \Big\{
    N(a\+ \mu\+ 1\+ t)+\uU-(a\+ 1)(\uuU-U) \Big\}
  \\
  & \times \Big\{
    (\uU-\dU\+ a\+ \mu)\big[ N\mu+(N\+a\+ \mu\+ 1)U-(N\+ a\+ \mu)\uU \big] 
  \\
  & \qquad\qquad
    +t\big[ \mu N-\mu\uU+(N\+ a\+ \mu)U-(N\+ a)\dU \big] \Big\}
  \\
  & \div \Big\{
    -\mu(a\+ \mu\+ t)+U-(N\+ a\+ \mu)(\uU-\dU) \Big\}
\end{split}
\label{U-adiff}
\end{equation}
where $ U := U_N(t;a,\mu), \dU := U_N(t;a\m 1,\mu), \uU := U_N(t;a\+ 1,\mu) $, 
etc and the boundary conditions are expressed by $ U_N(t;a,\mu) $ at three 
consecutive $ a $-values for all $ N, t, \mu $.
In addition $ U_N(t;a,\mu) $ satisfies a third order difference equation in 
$ \mu $ 
\begin{equation}
\begin{split}
  & -t(U\m \tU) = 
  \\
  & \phantom{div} \Big\{
    (\ttU\m U\+ 2N\+ a\+ \mu\+ 1)
    \big[ \m N(N\+ a)+(N\+ \mu\+ 1)U-(N\+ \mu)\tU \big]
    \\
  & \qquad\qquad
    -t\big[ \m N(N\+ a)+(\mu\+ 1)\ttU-(N\+ \mu\+ 1)\tU+NU \big] \Big\}
  \\
  & \div \Big\{
    -N(2N\+ a\+ \mu\+ 1\m t)+\tU-(N\+ \mu\+ 1)(\ttU-U) \Big\}
  \\
  & \times \Big\{
    (\tU\m \bU\+ 2N\+ a\+ \mu)
    \big[ \m N(N\+ a)+(N\+a\+ \mu\+ 1)U-(N\+ a\+ \mu)\tU \big]
    \\
  & \qquad\qquad
    -t\big[ \m N(N\+ a)-\mu\bU+(N\+ a\+ \mu)U-(N\+ a)\tU \big] \Big\}
  \\                                   
  & \div \Big\{
    -(N\+ a)(2N\+ a\+ \mu\m t)+U-(N\+ a\+ \mu)(\tU\m \bU) \Big\}
\end{split}
\label{U-mudiff}
\end{equation}
where now $ \bU := U_N(t;a,\mu\m 1), \tU := U_N(t;a,\mu\+ 1) $.
\end{prop}

\begin{proof}
The difference equation generated by $ T^{-1}_0 $ (\ref{dPIV-Ham}) with the
parameter identification $ v_2-v_1 = -\mu, v_3-v_1 = N+a, v_4-v_1 = N $
is not directly useful as this leads to a difference equation in both 
$N$ and $a$ (there are no difference equations in the parameter $N$ alone - 
only combined ones with $a$ or $\mu$). However difference equations generated
by the other shift operators can be simply found using this one with a
permutation of the parameter identification. Thus for the difference equation
in $a$ one can use $ v_2-v_1 = -\mu, v_3-v_1 = N, v_4-v_1 = N+a $ and 
the relation $ tH = U_N(t;a,\mu)+\mu N $ which gives the result is 
(\ref{U-adiff}). The second result (\ref{U-mudiff}) follows from the parameter 
identification $ v_2-v_1 = N, v_3-v_1 = N+a, v_4-v_1 = -\mu $ and 
$ tH = U_N(t;a,\mu)-N(N\+ a) $.
\end{proof}

Both of these difference equations are of the third order and linear in the
highest order difference and it may be possible to integrate these once and
reduce them to second order equations, however we do not pursue this 
question here. Although we have stated the difference equations for 
$ U_N(t;a,\mu) $, it is clear that $ V_N(t;a,\mu) $ satisfies these as well
although subject to different boundary conditions.

\subsection{$ \tilde{E}_N((0,s);a,\mu) $ for $ a \in \zz_{+} $}
In a previous study \cite{FH94} the quantities $\tilde{E}_N((0,s);a,0)$
and $\tilde{E}_N((0,s);a,2)$ for  $a \in \zz_{+}$ were expressed in terms
of $a \times a$ determinants. This was done using the method of orthogonal
polynomials to simplify the corresponding multiple integrals. In fact
we can easily express $\tilde{E}_N((0,s);a,\mu)$ for $a \in \zz_{+}$, with
$\mu$ general, as an $a \times a$ determinant using the methods of the
present study.

\begin{prop}
For $a \in \zz_{+}$ the function
\begin{align}\label{ff}
  \sigma^{(3)}(t) 
  & = t {d \over dt} \log \Big(
  t^{-N\mu-{1 \over 2}a(a-1)} e^{-(N+a)t} \det \Big[
     \delta^{j+k} ( e^t L_N^\mu(-t)) \Big]_{j,k=0,\dots,a-1} \Big)
  \nonumber \\
  & = t {d \over dt} \log \Big (
  t^{-N \mu} e^{-Nt} \det \Big[ {d^j \over dt^j}
         L_{N+k}^\mu(-t)  \Big]_{j,k=0,\dots,a-1} \Big )
\end{align}
Consequently, for $a \in \zz_{\ge 0}$,
\begin{equation}\label{ff0}
  \tilde{E}_N((0,s);a,\mu) \propto
  e^{-Ns} \det \Big[ {d^j \over ds^j} L_{N+k}^\mu(-s) \Big]_{j,k=0,\dots,a-1}.
\end{equation}
\end{prop}

\begin{proof}
Choose
\begin{equation}\label{ff1}
  v_3 \m v_1 = - N, \quad v_3 \m v_2 = \mu
\end{equation}
in (\ref{f1}). Then according to (\ref{det}), for $a \in \zz_{\ge 1}$
\begin{equation}\label{st}
  t^{-a/2} \bar{\tau}[a] = \det
  \Big [ \delta^{j+k} e^t {}_1 F_1(-N,\mu\+ 1;-t) \Big  ]_{j,k=0,\dots,a-1}.
\end{equation}
Using the fact that
\begin{equation*}
  {}_1 F_1(-N,\mu\+ 1;-t) \propto L_N^\mu(-t),
\end{equation*}
where $L_N^\mu$ denotes the Laguerre polynomial, substituting (\ref{st})
in (\ref{78'}) and then substituting the resulting expression in 
(\ref{6.5d})
gives the first formula for $\sigma^{(3)}$ in (\ref{ff}). In specifying
the parameters in (\ref{6.5d}) we have made use of (\ref{ff1}), the fact that
$v_4 - v_1 = a$, as well as (\ref{6.1a}). The theory noted in the sentence
containing (\ref{6.5c}) tells us that $\sigma^{(3)}$ satisfies (\ref{2.g})
with 
\begin{equation*}
  \nu_0 =0, \quad \nu_1=v_2 \m v_3=-\mu, \quad \nu_2=v_1 \m v_3 = N, \quad
  \nu_3 = v_4 \m v_3 = a \+ N.
\end{equation*}

To obtain the second equality in (\ref{ff}) we make use of the identity
\begin{equation*}
  \delta \left[ L_N^\mu(-t) e^t \right] =
   \left[ (N+1) L_{N+1}^\mu(-t) - (N+\mu+1)  L_N^\mu(-t) \right] e^t
\end{equation*}
in conjunction with elementary column operations, proceeding in an
analogous fashion to the derivation of (\ref{13.1}) using
the identity (\ref{12.1}). This shows 
\begin{equation*}
  \det \Big [
  \delta^{j+k} ( e^t L_N^\mu(-t)) \Big ]_{j,k=0,\dots,a-1} \propto
  \det \Big [
  \delta^j ( e^t L_{N+k}^\mu(-t)) \Big ]_{j,k=0,\dots,a-1}.
\end{equation*}
The second equality in (\ref{ff}) now follows by applying to this the
general identities
\begin{align*}
  \det \Big[ \delta^j (u(t) f_k(t)) \Big]_{j,k=0,\dots,a-1} 
  & = (u(t))^a \det  \Big[ \delta^j f_k(t)  \Big]_{j,k=0,\dots,a-1}
  \nonumber \\
  \det  \Big[ \delta^j f_k(t)  \Big]_{j,k=0,\dots,a-1} 
  & = t^{a(a-1)/2} \det \Big[ {d^j \over d t^j} f_k(t) \Big]_{j,k=0,\dots,a-1} .
\end{align*}

The function in the logarithm of the second equality in (\ref{ff})
is of the form $t^{-N\mu} e^{-Nt}$ times a polynomial in $t$. But for
$a \in \zz_{\ge 0}$, we know from the second integral formula in
(\ref{2.d}) that $t^{-N \mu} \tilde{E}((0,t);a,\mu)$ has the same
structure. Since $\sigma^{(3)}$ also satisfies the same differential equation
as $V_N(t;a,\mu)$ in (\ref{c2}), the formula (\ref{ff0}) follows.
\end{proof}

The determinant formula (\ref{st}) can be written as an $a$-dimensional
integral by using (\ref{89'}). Because (\ref{ff1}) implies an exponent
$-(N+a)$ for the factors $(1-u_j)$ in the integrand, we must first modify
the interval of integration. For this purpose it is convenient to first
change variables $u_j \mapsto 1 - u_j$. Then instead of the interval of
integration $[0,1]$ we choose a simple, closed contour which starts at
$u_j=1$ and encircles the origin. In particular, choosing this contour as 
the unit circle in the complex $u_j$ plane gives
\begin{equation*}
  t^{-a/2} \bar{\tau}[a] \propto t^{a(a-1)/2} e^{at}
  \Big \langle \prod_{j=1}^a(1+ e^{-2 \pi i x_j})^N (1 + e^{2 \pi i x_j})^\mu
  e^{t e^{2 \pi i x_j}} \Big \rangle_{{\rm  CUE}_a}.
\end{equation*}
Consequently we have the following generalization of (\ref{32a}).
\begin{prop}\label{p7}
For $a \in \zz_{\ge 0}$,
\begin{multline}\label{iI}
  \tilde{E}_N((0,s);a,\mu) \\
  = e^{-Ns} {M_a(0,0) \over M_a(\mu,N)}
  \Big\langle \prod_{j=1}^a(1+ e^{-2 \pi i x_j})^N (1 + e^{2 \pi i x_j})^\mu
  e^{s e^{2 \pi i x_j}} \Big\rangle_{{\rm  CUE}_a}.
\end{multline}
\end{prop}

We remark that the identity (\ref{iI}) is itself a special case of
a known more general integral identity \cite{Fo93c}. The latter
identity involves the PDFs
\begin{equation*}
  {1 \over C} \prod_{l=1}^N \lambda_l^a e^{-\beta \lambda_l / 2}
  \prod_{1 \le j < k \le N} |\lambda_k - \lambda_j |^\beta \quad
  (\lambda_l > 0)
\end{equation*}
\begin{equation*}
  {1 \over C} \prod_{1 \le j < k \le N} | e^{2 \pi i x_k} - e^{2 \pi i x_j}
  |^\beta \quad (-1/2 < x_l < 1/2)
\end{equation*}
defining what we will term the ensembles L$\beta$E and C$\beta$E
respectively. For $\beta = 2$, these PDFs were introduced in
(\ref{1.2}) and (\ref{1.29}) as the LUE and CUE. For $\beta = 1$ and
4 these PDFs also have a random matrix interpretation. They correspond
to the case of an orthogonal and symplectic symmetry respectively,
and give rise to the matrix ensembles LOE, COE and LSE, CSE.
Now define
\begin{equation*}
  \tilde{E}_N^{(\beta)}((0,s);a,\mu) =
  \Big \langle \prod_{l=1}^N \chi_{(s,\infty)}^{(l)} (\lambda_l -s)^\mu
  \Big \rangle_{{\rm L}\beta{\rm E}}
\end{equation*}
(note that this reduces to (\ref{2.d}) for $\beta = 2$). Then it
follows from results in \cite{Fo93c}, derived using the theory of
certain multi-variable hypergeometric functions based on Jack polynomials,
that for $a \in \zz_{\ge 0}$
\begin{multline}\label{iIa}
  \tilde{E}_N^{(\beta)}((0,s);a,\mu) 
  = e^{-Ns} {M_a^{(\beta)}(0,0) \over
             M_a^{(\beta)}(2(\mu\+1)/\beta \m 1,N)} \\
  \times
  \Big\langle \prod_{j=1}^a(1 + e^{2 \pi i x_j})^{2(\mu+1)/\beta - 1}
  (1 + e^{-2\pi i x_j})^N e^{s e^{2 \pi i x_j}} \Big\rangle_{{\rm C}(4/\beta)
  {\rm E}_a} ,
\end{multline}
where $M^{(\beta)}(a,b)$ denotes the integral (\ref{mn}) with exponent 2
in the product of differences replaced by $\beta$.

\section{$\tau$-function theory of \PIII and hard edge scaling}
\setcounter{equation}{0}
In the Laguerre ensemble the eigenvalues are restricted to be positive.
For $N$ large, the spacing between the eigenvalues in the neighbourhood of
the origin is of order $1/N$. By scaling the coordinates as in (\ref{1a})
the eigenvalue spacing is then of order 1 and well defined distributions
result. In this limit the \PV system degenerates to the \PIII system so it
is appropriate to revise the $\tau$-function theory of the latter.

\subsection{Okamoto $\tau$-function theory of \PIII }
Following \cite{Ok87a}, the Hamiltonian theory of \PIII 
(actually the \PIIIa system rather than the \PIII) 
can be formulated in terms of the Hamiltonian
\begin{equation}\label{H3}
  tH = q^2 p^2 - (q^2 + v_1q - t)p + {1 \over 2}(v_1 + v_2) q.
\end{equation}
Thus by substituting (\ref{H3}) in the Hamilton equations (\ref{6.1b}) 
and eliminating $p$ we find that $y(s)$ satisfies the \PIII differential 
equation
\begin{equation*}
  {d^2 y \over ds^2} = {1 \over y} \Big ( {dy \over ds} \Big )^2 -
  {1 \over s} {dy \over ds} + {1 \over s} (\alpha y^2 + \beta) +
  \gamma y^3 + {\delta \over y}
\end{equation*}
with $q(t)=sy(s)$, $t=s^2$ and
\begin{equation*}
  \alpha = - 4 v_2, \quad \beta = 4(v_1 + 1), \quad 
  \gamma = 4, \quad \delta = -4.
\end{equation*}
Note that with $H$ specified by (\ref{H3}), the first of the Hamilton
equations (\ref{6.1b}) gives
\begin{equation}
  tq' = 2q^2p - (q^2 + v_1 q - t).
\end{equation}
Thus, by using this equation to eliminate $p$ in (\ref{H3}), we see that
$tH$ can be expressed as an explicit rational function of $q$ and $q'$.
Analogous to Proposition \ref{p1}, it is straightforward to show that
$tH$ plus a certain linear function in $t$ satisfies a second order second
degree equation.

\begin{prop}\label{p1'}\cite{Ok87a}
With $H$ specified by (\ref{H3}), define the auxiliary Hamiltonian
\begin{equation}\label{H4}
  h = tH + {1 \over 4} v_1^2 -  {1 \over 2} t.
\end{equation}
The auxiliary Hamiltonian $h$ satisfies the differential equation
\begin{equation}\label{H5}
  (th'')^2 + v_1 v_2 h' - (4 (h')^2 - 1) (h - th') -
            {1 \over 4} (v_1^2 + v_2^2) = 0.
\end{equation}
\end{prop}

\begin{proof}
Following \cite{Ok87a}, we note from (\ref{H3}) and the
Hamiltonian equations  (\ref{6.1b}) that
\begin{align}\label{2.7a}
  h'   & = p - {1 \over 2} \nonumber \\
  th'' & = 2 (1 - p) pq + v_1 p -  {1 \over 2}(v_1 + v_2).
\end{align}
Using the first equation to substitute for $p$ in the second gives
\begin{equation*}
  q = {th'' - v_1 h' + {1 \over 2} v_2 \over
  {1 \over 2}(1 - 4 (h')^2)}, \qquad
  qp = {th'' - v_1 h' + {1 \over 2} v_2 \over (1 - 2h')},
\end{equation*}
while we can check from (\ref{H4}), (\ref{H3}) and the first
equation in (\ref{2.7a}) that
\begin{equation*}
  h - th' = (qp - {1 \over 2} v_1)^2 - q[qp - {1 \over 2}(v_1+v_2)].
\end{equation*}
Substituting for $qp$ and $q$, and simplifying, gives
(\ref{H5}).
\end{proof}

\medskip
Of interest in the random matrix application is the variant of (\ref{H5})
satisfied by
\begin{equation}\label{4.4'}
  \sigma_{III}(t) := - (tH) \Big |_{t \mapsto t/4} -
  {v_1 \over 4} (v_1 - v_2) + {t \over 4}.
\end{equation}
A straightforward calculation using the result of Proposition \ref{p1'}
shows that $\sigma_{III}$ satisfies
\begin{equation}\label{ss2}
  (t \sigma_{III}'')^2 - v_1 v_2 ( \sigma_{III}')^2 +
   \sigma_{III}' (4 \sigma_{III}' - 1) ( \sigma_{III} - t  \sigma_{III}')
  - {1 \over 4^3} (v_1 \m v_2)^2 = 0.
\end{equation}
Note that with the $\tau$-function defined in terms of the Hamiltonian
(\ref{H3}) by (\ref{2.11'}), we have
\begin{equation}\label{ss3}
  \sigma_{III}(t) = - t {d \over dt} \log \Big (
  e^{-t/4} t^{v_1 (v_1 - v_2)/4} \tau(t/4) \Big ).
\end{equation}

\subsection{B\"acklund transformations and Toda lattice equation}
For the Hamiltonian (\ref{H3}), Okamoto \cite{Ok87a} has identified two
B\"acklund
transformations with the property (\ref{u1}):
\begin{equation}\label{T}
  T_1\cdot\mathbf{v} = (v_1+1, v_2+1), \qquad
  T_2\cdot\mathbf{v} = (v_1+1, v_2 - 1).
\end{equation}
The operators $T_1$ and $T_2$ can be constructed out of more fundamental
operators $s_0, s_1, s_2$ associated with the underlying $B_2$ root
lattice, whose action (following \cite{Ok87a} and \cite{K99}) on $\mathbf{v},
p,q$ is given in Table \ref{t2}. According to Table \ref{t2} we have
\begin{equation}\label{T1}
  T_1 = s_0 s_2 s_1 s_2, \qquad
  T_2 = s_2s_0s_2s_1. 
\end{equation}

\begin{table}
\begin{center}
\begin{tabular}{|c||c|c|c|c|c|}\hline
 & $v_1$ & $v_2$ & $p$ & $q$ & $t$  \\ \hline
 $s_0$ & $-1\m v_2$ & $-1\m v_1$  & 
 ${\displaystyle q \over \displaystyle t}
   \left[q(p\m 1) - {1 \over 2}(v_1\m v_2)\right]\+ 1$ & 
 $-{\displaystyle t \over \displaystyle q} $ & $t$ \\
 $s_1$ & $v_2$ & $v_1$ &
 $p$ & $q + {\displaystyle v_2\m v_1 \over \displaystyle 2(p\m 1)} $ & $t$ \\
 $s_2$ & $v_1$  & $-v_2$ &  $1-p$ & $-q$ &
 $-t$   \\ \hline
\end{tabular}
\end{center}
\caption{\label{t2}  B\"acklund transformations relevant to the
\PIII Hamiltonian (\ref{H3}).}
\end{table}

Analogous to the situation with the Hamiltonian (\ref{tns}) in the \PV \\
theory, introducing the sequence of Hamiltonians by
\begin{equation}\label{Hs}
  H[n] := H \Big |_{(v_1,v_2) \mapsto (v_1+n,v_2+n)}
\end{equation}
where $H$ is given by (\ref{H3}), the operator $T_1$ can be used to
establish a Toda lattice equation for the corresponding $\tau$-function
sequence (\ref{tn}).

\begin{prop}
\cite{Ok87a,K99} The $\tau$-function sequence corresponding to the
Hamiltonian sequence (\ref{Hs}) obeys the Toda lattice equation
\begin{equation}\label{toda1j}
  \delta^2 \log \bar{\tau}[n] =
  {\bar{\tau}[n-1] \bar{\tau}[n+1] \over \bar{\tau}^2[n]}, \qquad
  \delta := t {d \over dt}
\end{equation}
where
\begin{equation}\label{toda2j}
  \bar{\tau}[n] := t^{n^2/2}  \tau[n](t/4).
\end{equation}
\end{prop}  

\begin{proof}
Analogous to (\ref{lhs}) and (\ref{A}) we have
\begin{equation}\label{lhsa}
  \delta \log {{\tau}[n-1] {\tau}[n+1] \over {\tau}^2[n]}
    = q[n](1-p[n])- T_1^{-1} q[n] (1 - T_1^{-1} p[n]) .
\end{equation}
Making use of (\ref{T1}) we can use Table \ref{t2} to explicitly
compute $ T_1^{-1} q[n]$ and $ T_1^{-1} p[n]$. This shows
\begin{multline*}
   q[n](1-p[n])- T_1^{-1} q[n] (1 - T_1^{-1} p[n]) \\
  = - {1 \over p[n]} \Big (2q[n] p^2[n] - (2q[n] + v_1)p[n]
                            + {1 \over 2}(v_1 + v_2) \Big ).
\end{multline*}
But according to (\ref{H3}) and the second equation in (\ref{2.7a}),
this latter expression is equal to $\delta \log({d \over dt}tH)$.
Substituting in (\ref{lhsa}) we deduce that
\begin{equation*}
  {d \over dt} \delta \log \tau[n] = C
  {\tau[n-1] {\tau}[n+1] \over {\tau}^2[n]}
\end{equation*}
Substituting for $ \tau[n]$ according to (\ref{toda2j}) and taking $C=1$
gives (\ref{toda1j}).
\end{proof}

\subsection{Classical solutions}
Okamoto \cite{Ok87a} has shown that for the special choice of
parameters $v_1 = - v_2$, the \PIII system admits a solution with
$\tau[0]=1$, and allows $\tau[1]$ to be evaluated as a Bessel
function. 

\begin{prop}\label{PIII_class} \cite{Ok87a} For the special choice of 
parameters
\begin{equation}\label{v1.9}
  v_1 = - v_2
\end{equation}
in (\ref{H3}) it is possible to choose $\tau[0]=1$. The first member
$\tau[1]$ of the $\tau$-function sequence corresponding to (\ref{Hs}),
after the substitution $t \mapsto t/4$, satisfies the equation
\begin{equation}\label{v20a}
  t (\hat{\tau}[1])'' + (v_1 + 1)  (\hat{\tau}[1])' - {1 \over 4}
  \hat{\tau}[1] = 0, \quad \hat{\tau}[1] := \tau[1](t/4).
\end{equation}
For $v_1 \notin \zz$, this equation has the two linearly independent
solutions in terms of Bessel functions
\begin{equation}\label{v20}
  \hat{\tau}[1] = t^{-v_1/2} I_{\pm v_1}(\sqrt{t})
\end{equation}
(for $v_1 \in \zz$, $I_{v_1}$ and $I_{-v_1}$ are proportional and
(\ref{v20}) only provides one independent solution).
\end{prop}

\begin{proof}
Substituting (\ref{v1.9}) into the definition (\ref{H3})
of $H$ we see that it is possible to choose
\begin{equation}
  p=0, \quad H=0,
\end{equation}
the latter allowing us to take $\tau[0]=1$. To calculate $q$ we use the
Hamilton equation
\begin{equation}\label{su}
  t q' = {\partial t H \over \partial p} \Big |_{v_1 = -v_2 \atop
  p=0} = - (q^2 + v_1 q - t).
\end{equation}
Now, it follows from (\ref{TH}) with $H$ given by (\ref{H3}) and
$T^{-1}_{0}$ by $T_1$ that
\begin{equation*}
  q = t {d \over dt}\log \tau[1].
\end{equation*}
Substituting this in (\ref{su}) and changing variables $t \mapsto
t/4$ gives (\ref{v20a}).
\end{proof}

For definiteness take the $+$ sign in (\ref{v20}) and let $v_1 =
\nu$. This substituted in (\ref{toda2j}) with $n=1$, and the corresponding
formula for $\bar{\tau}[1]$ substituted in (\ref{det}) shows that
\begin{equation}\label{detI}
  t^{n(\nu-1)/2} \bar{\tau}[n] = \det [ \delta^{j+k} I_\nu(\sqrt{t})
  ]_{j,k=0,\dots,n-1},
\end{equation}
where use has also been made of the theory in the first sentence of the
proof of Proposition \ref{pU}. The determinant in (\ref{detI}) can be
written in a form which is independent of the operator $\delta$.

\begin{prop}\label{p6}
We have
\begin{equation}\label{dk0}
   \det [ \delta^{j+k} I_\nu(\sqrt{t})
  ]_{j,k=0,\dots,n-1} = (t/4)^{n(n-1)/2} \det [I_{j-k+ \nu}(\sqrt{t})
  ]_{j,k=0,\dots,n-1}.
\end{equation}
\end{prop}

\begin{proof}
Let $t = s^2$ and note that
\begin{equation*}
  \delta := t {d \over dt} = {1 \over 2} s {d \over ds} =: {1 \over 2}
  \delta_s
\end{equation*}
to conclude
\begin{equation}\label{dk}
  \det [ \delta^{j+k} I_\nu(\sqrt{t})
  ]_{j,k=0,\dots,n-1} = 2^{-n(n-1)} \det [ \delta_s^{j+k} I_\nu(s)
  ]_{j,k=0,\dots,n-1}.
\end{equation}
We now adopt a very similar strategy to that used in the proofs of
Propositions \ref{pU} and \ref{p4}. Briefly, we first make use of
the identity
\begin{equation*}
  \delta_s I_\nu(s) = s I_{\nu+1}(s) + \nu I_\nu(s)
\end{equation*}
together with elementary row operations to eliminate the operator
$\delta^j$ in (\ref{dk}), obtaining
\begin{equation}\label{dk1}
  \det [ \delta_s^{j+k} I_\nu(s)
  ]_{j,k=0,\dots,n-1} = \det [ \delta_s^k s^{j} I_{\nu + j}(s)
  ]_{j,k=0,\dots,n-1}.
\end{equation}
Next, use is made of the identity
\begin{equation}\label{id1}
  \delta_s s^j I_{\nu+j}(s) = s^{j+1} I_{\nu+j-1}(s) - \nu 
  s^j I_{\nu+j}(s)
\end{equation}
to perform elementary column operations thus reducing the RHS of (\ref{dk1})
to the form
\begin{equation}\label{dk2}
  \det [s^j I_{\nu+j}(s) \;\ldots\; \delta_s^{k-1} s^{j+1} I_{\nu + j-1}(s)
        \;\ldots ]_{j = 0,\dots,n-1 \atop k=1,\dots,n-1}.
\end{equation}
Next we apply elementary column operations, using the identity
(\ref{id1}) with $j \mapsto j+1$, $\nu \mapsto \nu - 2$, to show that
(\ref{dk2}) is equal to
\begin{equation*}
  \det [s^j I_{\nu+j}(s) \;\; s^{j+1} I_{\nu+j-1}(s)
  \;\ldots\; \delta_s^{k-2} s^{j+2} I_{\nu + j-2}(s) \;\ldots
  ]_{j = 0,\dots,n-1 \atop k=2,\dots,n-1}
\end{equation*}
Continuing in this fashion, using the identity  (\ref{id1}) with appropriate
substitutions, 
reduces the RHS of (\ref{dk1}) to
\begin{equation*}
  \det [s^{j+k}  I_{\nu+j -k}(s) ]_{j,k=0,\dots,n-1}.
\end{equation*}
Substituting in (\ref{dk}) gives (\ref{dk0}).
\end{proof}

\begin{cor} \label{cr1} 
The function
\begin{equation}\label{prvj}
  \sigma_{III}(t) = - t {d \over dt} \log \Big ( e^{-t/4} t^{\nu^2/2}
  \det[ I_{j-k+\nu}(\sqrt{t}) ]_{j,k=0,\dots,n-1} \Big )
\end{equation}
satisfies the equation (\ref{ss2}) with parameters
\begin{equation}\label{prv}
  (v_1, v_2) = (\nu+n, -\nu+n)
\end{equation}
and boundary condition
\begin{equation}\label{prv1}
  \sigma_{III}(t) \mathop{\sim}\limits_{t \to \infty}
  {t \over 4} - {n t^{1/2} \over 2} - \Big ( {\nu^2 \over 2} - {n^2 \over 4}
  \Big ) + \cdots 
\end{equation}
\end{cor}

\begin{proof}
It follows from Proposition \ref{p6}, (\ref{detI}) and (\ref{toda2j}) that
\begin{equation*}
  \tau[n](t/4) \propto 
   t^{-n\nu/2} \det [ I_{j-k+\nu}(\sqrt{t}) ]_{j,k=0,\dots,n-1}\end{equation*}
The equation satisfied by $\sigma_{III}(t)$
follows by substituting this, and the parameter
values (\ref{prv}), in (\ref{ss3}). To obtain the boundary condition, we
make use of the well known Toeplitz determinant formula
\begin{multline*}
  \det\Big[ {1 \over 2 \pi} \int_{-\pi}^\pi f(\theta) e^{-i(j-k)\theta}
  \, d\theta \Big]_{j,k=0,\dots,n-1} \\
  = {1 \over n!} {1 \over (2 \pi)^n}
  \prod^{n}_{l=1} \int_{-\pi}^\pi d \theta_{l} \, f(\theta_{l})
  \prod_{1 \le j < k \le n} | e^{i \theta_k} - e^{i \theta_j} |^2
\end{multline*}
together with the integral representation
\begin{equation*}
  I_n(z) = {1 \over 2 \pi} \int_{-\pi}^\pi e^{-in \theta + z \cos \theta}
  \, d \theta, \qquad n \in \zz
\end{equation*}
to rewrite the determinant in (\ref{prvj}) as a multidimensional integral,
\begin{multline}\label{plo}
  \det[ I_{j-k+\nu}(\sqrt{t}) ]_{j,k=0,\dots,n-1} \\
  = {1 \over n!} {1 \over (2\pi)^n}
  \prod^{n}_{l=1} \int_{-\pi}^\pi d\theta_{l} \,
  e^{\sqrt{t} \cos\theta_{l}-i\nu\theta_{l}}
  \prod_{1 \le j < k \le n} | e^{i \theta_k} - e^{i \theta_j} |^2
\end{multline}
valid for $\nu \in \zz$.
For large $t$ the dominant contribution to the above integral comes from
the neighbourhood of $\theta_j=0$ ($j=1,\dots,n)$. Expanding to leading
order about these points and changing variables shows
\begin{equation*}
  \det[ I_{j-k+\nu}(\sqrt{t}) ]_{j,k=0,\dots,n-1}
  \mathop{\sim}\limits_{t \to \infty}
  C e^{n \sqrt{t}} t^{-n^2/4}.
\end{equation*}
Substituting this in (\ref{prvj}) gives (\ref{prv1}).
\end{proof}
 
The result of Corollary \ref{cr1} is relevant to the hard edge scaling of
$\tilde{E}_N((0,s);a,\mu)$, which gives the quantity $\tilde{E}^{\rm hard}
(t;a,\mu)$ defined by (\ref{1.54}). 
Now, in an earlier study \cite{FH94}, it was shown that for $a \in \zz_{\ge
0}$
\begin{align}\label{jm}
  \tilde{E}^{\rm hard}(t;a,0) 
  & = e^{-t/4} \det[ I_{j-k}(\sqrt{t}) ]_{j,k=0,\dots,a-1}
  \nonumber\\
  \tilde{E}^{\rm hard}(t;a,2)
  & \propto e^{-t/4} t^{-a} \det[ I_{j-k+2}(\sqrt{t}) ]_{j,k=0,\dots,a-1} 
\end{align}
Thus, as already noted in 
\cite{FH94} in the case of
$\tilde{E}^{\rm hard}(t;a,0)$ (deduced from knowledge of (\ref{1.55})), it
follows from Corollary \ref{cr1} that $\tilde{E}^{\rm hard}(t;a,\mu)$
for $\mu=0$ and $\mu=2$ can be characterised as the solution of the 
equation (\ref{ss2}) with parameters $(v_1,v_2) = (a+\mu, a- \mu)$. 

The results (\ref{jm}) were obtained by computing the limit of the RHS of
(\ref{ff0}) using the asymptotic formula
\begin{equation*}
  e^{-x/2} x^{a/2} L_N^a(-x) \sim N^{a/2} I_a(2(Nx)^{1/2}).
\end{equation*}
The same approach allows $\tilde{E}^{\rm hard}
(t;a,\mu)$ to be computed for general $\mu$ and $a \in \zz_{\ge 0}$,
giving the result
\begin{equation}\label{pl}
  \tilde{E}^{\rm hard}(t;a,\mu) \propto
  e^{-t/4} t^{-\mu a/2}\det [ I_{j-k+\mu}(\sqrt{t}) ]_{j,k=0,\dots,a-1}.
\end{equation}
It then follows from Corollary \ref{cr1} that the result
(\ref{sa2}) holds for general $\mu$ and 
$a \in \zz_{\ge 0}$. For general $a > -1$ the
result (\ref{sa2}) can be deduced from the first formula in (\ref{32}).
This task will be undertaken in subsection 4.5.

To conclude this subsection we make two remarks. The first is
that substituting (\ref{plo}) in (\ref{pl}) shows that for $\mu \in \zz$ 
and $a \in \zz_{\ge 0}$
\begin{align}\label{4.31}
  \tilde{E}^{\rm hard}(t;a,\mu)
  & \propto e^{-t/4} t^{-\mu a/2}
    \prod^{a}_{l=1} \int_{-\pi}^\pi d \theta_{l} \,
    e^{\sqrt{t}\cos\theta_{l}-i\mu\theta_{l}}
    \prod_{1 \le j < k \le a} | e^{i\theta_k} - e^{i\theta_j} |^2 
  \nonumber \\
  & \propto e^{-t/4} t^{-\mu a/2} \Big \langle 
    e^{{1 \over 2} \sqrt{t} {\rm Tr} (U + \bar{U})}
    (\det U)^{-\mu} \Big \rangle_{U \in {\rm CUE}_a}.
\end{align}
This identity, relating an average in the infinite LUE
scaled at the hard edge to an average in the CUE of dimension $a$,
has previously been derived \cite{Fo93c} as a scaled limit of (\ref{iI}).
With $c := 2(\mu+1)/\beta$ and $\mu$ such that $c \in \zz_{\ge 0}$,
it was also shown that the identity (\ref{iIa}) in the hard edge
limit reduces to
\begin{multline}\label{4.32}
  \tilde{E}^{(\beta){\rm hard}}(t;a,\mu) 
   = C e^{-t/4} \Big ( {4 \over t} \Big )^{(c-1)a}
  \Big ( {1 \over 2 \pi} \Big )^a
  \\ \times
  \prod^{a}_{l=1} \int_{-\pi}^\pi d \theta_{l} \,
  e^{\sqrt{t}\cos\theta_{l}-i(c-1)\theta_{l}}
  \prod_{1 \le j < k \le a} | e^{i\theta_k} - e^{i\theta_j} |^{4/\beta}
\end{multline}
where
\begin{equation*}
  C = \prod_{j=1}^a {\Gamma(1+2/\beta) \Gamma(c+2(j-1)/\beta) \over
  \Gamma(1 + 2j/\beta)}
\end{equation*}
and $\tilde{E}^{(\beta){\rm hard}}$, like $\tilde{E}^{\rm hard}$ specified
by (\ref{1.54}), is normalised so that it equals unity for $t=0$.
As a second remark, we note that a feature of the differential
equation (\ref{H5}) is that it is unchanged by the mapping
$t \mapsto - t$, $v_2 \mapsto - v_2$. Using this, it follows
from (\ref{H4}) and (\ref{4.4'}) that (\ref{ss2}) is also satisfied by
\begin{equation}\label{n.2}
  \sigma_{III}(t) \Big |_{v_2 \mapsto - v_2 \atop
  t \mapsto - t} \+ {t \over 2} \+ v_1 v_2/2 =
  - t {d \over dt} \log \Big ( e^{-t/4} t^{v_1(v_1-v_2)/4}
  \tau(-t/4) \Big |_{v_2 \mapsto - v_2} \Big )
\end{equation}
where the equality follows from (\ref{ss3}). But by an appropriate
modification of Propositions \ref{PIII_class} and \ref{p6}, a determinant
formula for $\tau(-t/4) \Big |_{v_2 \mapsto - v_2}$ can be given
for certain $(v_1,v_2)$. Performing these
modifications and substituting the resulting formula in (\ref{n.2})
leads us to the conclusion that
\begin{equation}\label{n.3}
  -t {d \over dt} \log \Big ( e^{-t/4} t^{\nu^2/2}
  \det [ J_{j-k+\nu}(\sqrt{t}) ]_{j,k=0,\dots,n-1} \Big )
\end{equation}
satisfies (\ref{ss2}) with parameters
\begin{equation*}
  (v_1,v_2) = (\nu+n,\nu-n).
\end{equation*}
In the $n=\mu=2$ and $ \nu=a $ case, substituting (\ref{n.3}) for $\sigma(-t)$ 
in (\ref{1.73a}) gives the well known \cite{Fo93a} expression for
$\rho^{\rm hard}(s)$ in terms of Bessel functions.

\subsection{Schlesinger Transformations}
Integrable difference equations are found to arise from the Schlesinger
transformations generated by $ T_1, T_2 $ 
\cite{ROG-2000,RGTT-2000,NSKGR-96} acting on $ q $ and $ p $ and we consider 
these here, along with the difference equations for the Hamiltonians.

\begin{prop}\cite{RGTT-2000}
The Schlesinger transformations of the \PIII system for the operators 
$ T_1 (T_2) $ generating the parameter sequences $ (v_1\+ n,v_2\+ n) $
(respectively $ (v_1\+ n,v_2\m n) $) acting on the transcendent $ q $ 
correspond to
second order difference equations of the alternate discrete Painlev\'e II, 
d-\PII, equation (respectively its dual)
\begin{align}
   \frac{1}{2}{v_1+v_2+2+2n \over q[n]q[n+1]+t}
 + \frac{1}{2}{v_1+v_2+2n \over q[n-1]q[n]+t} 
 & = q^{-1}-{q \over t}+{v_2+n \over t} , 
 \label{d-PII:a} \\
   \frac{1}{2}{v_2-v_1-2-2n \over q[n]q[n+1]-t}
 + \frac{1}{2}{v_2-v_1-2n \over q[n-1]q[n]-t} 
 & = q^{-1}+{q \over t}+{v_1+n \over t} .
 \label{d-PII:b}
\end{align}
Further, under the action of $ T_1 $, $ tH[n+1]-tH[n] = q[n](1-p[n]) $
and this satisfies the second order difference equation
\begin{equation}
\begin{split}
   q(1\m p)[n]
  & \Big\{ q(1\m p)[n]+\frac{1}{2}(v_1-v_2) \Big\}
  \\
  & = t 
    \Big\{ q(1\m p)[n+1]+q(1\m p)[n]+\frac{1}{2}(v_1-v_2) \Big\}
  \\
  & \div
    \Big\{ q(1\m p)[n+1]+q(1\m p)[n]+v_1+1+n \Big\}
  \\
  & \times
    \Big\{ q(1\m p)[n]+q(1\m p)[n-1]+\frac{1}{2}(v_1-v_2) \Big\}
  \\
  & \div
    \Big\{ q(1\m p)[n]+q(1\m p)[n-1]+v_1+n \Big\} ,
\end{split}
\label{III-T1diff}
\end{equation}
while under the action of $ T_2 $, $ tH[n+1]-tH[n] = q[n]p[n] $ and this 
satisfies the second order difference equation
\begin{equation}
\begin{split}
   qp[n]\Big\{ qp[n]-\frac{1}{2}(v_1+v_2) \Big\} = 
  & -t
    \Big\{ qp[n+1]+qp[n]-\frac{1}{2}(v_1+v_2) \Big\}
  \\
  & \div
     \Big\{ qp[n+1]+qp[n]-v_1-1-n \Big\}
  \\
  & \times
     \Big\{ qp[n]+qp[n-1]-\frac{1}{2}(v_1+v_2) \Big\}
  \\
  & \div
     \Big\{ qp[n]+qp[n-1]-v_1-n \Big\} .
\end{split}
\label{III-T2diff}
\end{equation}
\end{prop}

\begin{proof}
In the case of the $ T_1 $ transformation the action on the canonical 
variables in the forward and reverse directions are
\begin{align}
   q[n+1] & = -{t \over q}
            + \cfrac{\frac{1}{2}(v_1+v_2+2+2n)t}
                    {q[q(p-1)-\frac{1}{2}(v_1-v_2)]+t}
          \label{XfmIII:a}  \\
   p[n+1] & = {q \over t}[q(p-1)-\frac{1}{2}(v_1-v_2)]+1 
          \label{XfmIII:b} \\
   q[n-1] & = \cfrac{t}{\cfrac{\frac{1}{2}(v_1+v_2+2n)}{p}-q} 
          \label{XfmIII:c} \\
   p[n-1] & = 1
            - t^{-1}\left[ \cfrac{\frac{1}{2}(v_1+v_2+2n)}{p}-q \right]
          \label{XfmIII:d}\\
          & \qquad \times
   \left\{ \left[ \cfrac{\frac{1}{2}(v_1+v_2+2n)}{p}-q \right](1-p)
           -\frac{1}{2}(v_1-v_2) \right\}
          \nonumber
\end{align}
By combining (\ref{XfmIII:a}) and (\ref{XfmIII:b}) into the form
\begin{equation}
   q[n+1] + {t \over q[n]} = {\frac{1}{2}(v_1+v_2+2+2n) \over p[n+1]}
\end{equation}
and eliminating $ p $ between this equation and (\ref{XfmIII:c}) we have
the result (\ref{d-PII:a}). The difference equation for the Hamiltonian can 
be found by solving 
\begin{multline}
  q(1\m p)[n+1] + q(1\m p)[n] +\frac{1}{2}(v_1-v_2) \\
  = -\frac{1}{2}(v_1+v_2+2+2n)
     { q(1\m p)[n] +\frac{1}{2}(v_1-v_2) \over 
       q(1\m p)[n] +\frac{1}{2}(v_1-v_2) -t/q }
\end{multline}
for $ q $ and
\begin{equation}
 q(1\m p)[n-1] + q(1\m p)[n] +\frac{1}{2}(v_1-v_2) 
  = \frac{1}{2}(v_1+v_2+2n){1-p[n] \over p[n]}
\end{equation}
for $ p $ and then reforming $ q(1-p) $. The result is (\ref{III-T1diff}).
The corresponding results for the $ T_2 $ sequence can be found in a similar
manner. 
\end{proof}

\subsection{Hard edge scaling $\tilde{E}^{\rm hard}(t;a,\mu)$}
According to (\ref{1.54}) and (\ref{32}) we have for general
$a > - 1$
\begin{equation}\label{8.1}
  \tilde{E}^{\rm hard}(s;a,\mu) = \exp \int_0^s v(t;a,\mu) \, {dt \over t}
\end{equation}
where
\begin{equation}\label{8.2}
  v(t;a,\mu) = \lim_{N \to \infty} \Big( V_N(t/4N;a,\mu) + \mu N \Big).
\end{equation}
In this limit the differential equation (\ref{2.g}) characterizing the \PV
auxiliary Hamiltonian (\ref{6.5a}) and specifying $V_N$
degenerates to the differential equation (\ref{H5}) characterizing
the \PIII auxiliary Hamiltonian (\ref{H4}) thus identifying $v$ with
this quantity.

\begin{prop}\label{p12}
The function $v$ specified by (\ref{8.2}) satisfies the differential equation
\begin{equation}\label{8.3}
(tv'')^2 - (\mu + a)^2(v')^2 - v'(4 v' + 1)(v - t v')
- {\mu \over 2}(\mu+a) v' - {\mu^2 \over 4^2} = 0.
\end{equation}
Consequently
\begin{equation}\label{8.4}
v(t;a,\mu) = - \Big ( \sigma_{ III}(t) + \mu(\mu+a)/2 \Big )
\end{equation}
where $ \sigma_{ III}$ is specified by the Jimbo-Miwa-Okamoto $\sigma$-form
of the Painlev\'e III equation (\ref{sa}) with parameters (\ref{sa3}) 
$ v_1 = a+\mu, v_2 = a-\mu $ and is subject to the boundary condition 
(\ref{sa4})
\begin{equation}
   v(t;a,\mu) \mathop{\sim}\limits_{t \to \infty}
   - \frac{1}{4}t + \frac{1}{2}at^{1/2} - \frac{1}{4}a(a+2\mu) .
\end{equation}
\end{prop}

\begin{proof}
Making the replacement $\sigma \mapsto \sigma - N \mu$
in (\ref{2.g}) with parameters (\ref{pv}), changing variables
$t \mapsto t/4N$, $\sigma(t/4N) \mapsto v(t)$, and equating
terms of order $N^2$ (which is the leading order) on both sides
gives (\ref{8.3}). Substituting (\ref{8.4}) in (\ref{8.3})
shows $\sigma_{III}$ satisfies (\ref{sa}) with parameters as
specified. The boundary condition is the same as  (\ref{prv1})
with $n=a$, $\nu = \mu$.
\end{proof}

\begin{prop}\label{HE_diff}
The function $ v(t;a,\mu) $ satisfies a third order difference equation in
the variable $ a $ (suppressing the additional dependencies)
\begin{multline}
   \frac{1}{4}a(a+1)t =
   [v(a\+ 1)-v(a\m 1)\+ a\+ \mu][v(a\+ 2)-v(a)\+ a\+ \mu\+ 1] \\
   \times
   \left[ \frac{1}{4}t-av(a\+ 1)+(a\+ 1)v(a) \right] ,
\label{HE-adiff}
\end{multline}
and a third order difference equation in $ \mu $
\begin{multline}
  -\frac{1}{4}t =
   \left[ v(\mu\+ 1)-v(\mu) \right]\left[ v(\mu\+ 1)-v(\mu)+a \right] \\
   \times
   { [v(\mu\+ 1)-v(\mu\m 1)\+ a\+ \mu][v(\mu\+ 2)-v(\mu)\+ a\+ \mu\+ 1]
   \over 
     [v(\mu\+ 1)-v(\mu\m 1)\+ a][v(\mu\+ 2)-v(\mu)\+ a] }
\label{HE-mudiff}
\end{multline}
\end{prop}

\begin{proof}
These two results follow from applying the hard edge scaling form
(\ref{8.2}) to the finite-$N$ difference equations (\ref{U-adiff}) and
(\ref{U-mudiff}). Note that (\ref{HE-mudiff}) is precisely the result 
that would be inferred from the \PIII difference equation generated by
the $ T_2 $ shift (\ref{III-T2diff}) with the correspondence
\begin{equation}
   v(t;a,\mu) = tH\Big|_{t \mapsto t/4} - t/4 .
\end{equation}
\end{proof}

We have noted in (\ref{rt1}) that a corollary of Proposition \ref{p12}
is the evaluation of $p_{\rm min}(s;a)$ in terms of $v(t;a,2)$. Since,
analogous to (\ref{R0}),
\begin{equation}
  p_{\rm min}^{\rm hard}(s;a)  =  -
  {d \over ds} \tilde{E}^{\rm hard}((0,s);a,0) ,
\end{equation}
the results (\ref{rt1}) and (\ref{1.55}) substituted into this formula
imply an identity between transcendents analogous to (\ref{i1}). As is
the case with (\ref{i1}), this identity too can be independently verified.

\begin{prop}
With $\sigma(t)$ denoting the auxiliary Hamiltonian (\ref{4.4'}) with
parameters $(v_1,v_2) = (a+2,a-2)$, and $\sigma_B(t)$ denoting the same
quantity with parameters $(v_1, v_2) = (a,a)$, the identity
\begin{equation}\label{sw}
  \sigma(t) = \sigma_B(t) - 1 - t { \sigma_B'(t) \over \sigma_B(t) },
\end{equation}
holds.
\end{prop}

\begin{proof}
The indirect derivation of this result has been sketched above. A direct
derivation can be given by using the properties of the shift operator
$T_2$ in (\ref{T}) and (\ref{T1}). First, denote by $t H[0]$ the
Hamiltonian $tH$ in (\ref{4.4'}) (and similarly define $p[0]$, $q[0]$)
so that
\begin{equation}\label{pa2'}
  \sigma_B(t) = - (tH[0]) \Big |_{t \mapsto t/4} + t/4.
\end{equation}
It then follows from the definitions of $T_2$ and $\sigma(s)$ that
\begin{equation}\label{pa2}
  \sigma(t) = - (t T_2^2 H[0]) \Big |_{t \mapsto t/4} - (a+2) +  t/4.
\end{equation}
On the other hand, from the definition (\ref{H3}) and the property
(\ref{u1}) of $T_2$ we have that
\begin{equation*}
  t T_2^2 H[0] = t  H[0] - q[0] p[0] - (T_2 q[0])(T_2 p[0]).
\end{equation*}
But from (\ref{T1}) and Table \ref{t2} it follows that
\begin{align*}
  T_2 q[0] 
  & = \Big ( {t \over q[0]} - {t \over q[0](q[0] p[0] - a) + t} \Big ) \\
  T_2 p[0] 
  & = - { q[0] \over t} (q[0] p[0] - a)
\end{align*}
and thus
\begin{equation}\label{pa3}
  t T_2^2 H[0] = t  H[0] - a - 1 + t {\sigma_B'(t) \over
  \sigma_B(t)}
\end{equation}
where to obtain the final term use has been made of (\ref{H3}),
(\ref{4.4'}) and the analogue of the first equation in (\ref{2.7a}).
Substituting in (\ref{pa2}) and recalling (\ref{pa2'}) gives (\ref{sw}).
\end{proof}

\subsection{The spacing probability $p_2(0;s)$}
The evaluation (\ref{8.1}) has  consequence regarding the nearest
neighbour spacing distribution $p_2(0;s)$ for the scaled, infinite GUE
in the bulk. This quantity has a special place in the Painlev\'e
transcendent evaluation of gap probabilities because it motivated
the study of the probability  $E_2(0;s)$ of
no eigenvalues in an interval of length $s$
in the scaled, infinite GUE through the relation
\begin{equation*}
  p_2(0;s) = {d^2 \over d s^2} E_2(0;s),
\end{equation*}
and $E_2(0;s)$ in turn was the first quantity in random matrix theory to
be characterised as the solution of a non-linear equation in the Painlev\'e
theory \cite{JMMS80}. More recently, building on the evaluation of
$E_2(0;s)$ from  \cite{JMMS80}, it has been shown that \cite{FW99}
\begin{equation}\label{fw3}
  p_2(0;s) = - {\tilde{\sigma}(\pi s) \over s}
\exp \int_0^{\pi s} { \tilde{\sigma}(t) \over t} \, dt
\end{equation}
where $ \tilde{\sigma}(s) $ is specified by the solution of the non-linear
equation
\begin{equation*}
  s^2 (\tilde{\sigma}'')^2 + 4(s \tilde{\sigma}' - \tilde{\sigma})
  (s  \tilde{\sigma}' - \tilde{\sigma} + ( \tilde{\sigma}')^2) -
  4  ( \tilde{\sigma}')^2 = 0
\end{equation*}
(a close relative of (\ref{2.g}) for a particular choice of the parameters)
subject to the boundary condition $ \tilde{\sigma}(s) \sim - (s^3/3 \pi)$.
Here we will provide a Painlev\'e transcendent evaluation of 
$p_2(0;s)$ distinct from (\ref{fw3}).

The starting point is the identity
\begin{multline}\label{hh}
  \Big \langle \prod_{l=1}^N (\lambda_l^2 - s^2)^2 
  \chi_{(-\infty,-s) \cup (s,\infty)}^{(l)} \Big \rangle_{\rm GUE} \\
  \propto
  \tilde{E}_{[N/2]}((0,s^2); {1 \over 2}, 2)
  \tilde{E}_{[(N+1)/2]}((0,s^2); - {1 \over 2}, 2),
\end{multline}
which is a variant of a formula given in \cite{Fo99a}, applicable to
general matrix ensembles with a unitary symmetry and an even weight function.
Now the LHS of (\ref{hh}) multiplied by $s^2$ is proportional to the
density function for a spacing between eigenvalues of length $2s$
symmetric about the origin. Recalling that the bulk scaling in the GUE
requires
\begin{equation*}
  \lambda_l \mapsto {\pi \lambda_l \over \sqrt{2 N}}
\end{equation*}
it follows by replacing $s$ by $\pi s / \sqrt{2N}$, making use of the
definition (\ref{1.54}), and the property 
$p_2(0;s) \sim {\pi^2 \over 3} s^2$ which follows from the
boundary condition for $\tilde{\sigma}(s)$ in
(\ref{fw3}) (this fixes the proportionality
constant) that
\begin{equation}
  p_2(0;2s) = {(2\pi s)^2 \over 3}
  \tilde{E}^{\rm hard}((\pi s)^2; {1 \over 2}, 2)
  \tilde{E}^{\rm hard}((\pi s)^2; - {1 \over 2}, 2).
\end{equation}
According to (\ref{8.1}) this specifies $p_2(0;s)$ in terms of 
Painlev\'e III transcendents whereas (\ref{fw3}) involves
Painlev\'e V transcendents.

\section{Concluding remarks}
\setcounter{equation}{0}
The results of the paper have already been summarised in Section 1.
Here we want to draw attention to a feature of this work (and our previous
work \cite{FW00}) which requires further study. This feature  relates to
the boundary conditions in the scaled limits, in particular the specification
of $\tilde{E}^{\rm hard}(s;a,\mu)$ by (\ref{sa2}). One observes that the
boundary condition (\ref{sa4}) is even in $\mu$, as is the differential
equation (\ref{sa}) with parameters (\ref{sa3}). Thus according to this
specification the solution itself must be even in $\mu$, but this
contradicts the small $t$ behaviour which for the formula
(\ref{sa2}) to be well defined must be
\begin{equation}\label{6.m}
   \sigma(t) \mathop{\sim}\limits_{t \to 0} - \frac{1}{2}\mu(\mu + a)
   + O(t^{\epsilon}), \quad (\epsilon > 0) .
\end{equation}

The situation is well illustrated by the case $a=1$, for which
(\ref{prvj}) shows 
\begin{equation}
  \sigma(t) = - t {d \over dt} \log \Big ( e^{-t/4} t^{\mu^2/2}
  I_\mu(\sqrt{t}) \Big ).
\end{equation}
The small $t$ expansion of $I_\mu(\sqrt{t})$ shows
\begin{equation}
  \sigma(t) \mathop{\sim}\limits_{t \to 0} - \frac{1}{2}\mu(\mu + 1) +
  {\mu \over 4 (\mu + 1)} t,
\end{equation}
in agreement with (\ref{6.m}), whereas the large $t$ expansion of
$I_\mu(\sqrt{t})$ shows
\begin{equation}
  \sigma(t) \mathop{\sim}\limits_{t \to \infty} {t \over 4} -
  {t^{1/2} \over 2} + \sum_{l=0}^\infty {a_l \over t^{l/2}}
\end{equation}
where the $a_l$ are even functions of $\mu$. Thus the asymmetry between
$\mu$ and $-\mu$ can only be present in an exponentially small term for
$t \to \infty$. This is clear from the exact expression
\begin{equation}
   \sigma(t;\mu)-\sigma(t;-\mu) = 
   -{\sin\pi\mu \over \pi}{1 \over I_{\mu}(\sqrt{t})I_{-\mu}(\sqrt{t})} .
\end{equation}
Furthermore, although the boundary condition
(\ref{6.m}) distinguishes the cases $\pm \mu$, the differential equation
(\ref{sa}) also requires that the leading (in general) non-analytic
term also be specified for the solution to be uniquely specified
(for example, with $\mu = 0$ this term is proportional to $t^{a+1}$).

A practical consideration of this discussion is that the boundary
condition (\ref{sa4}) does not uniquely determine the solution of
(\ref{sa}), so that $\tilde{E}^{\rm hard}(s;a,\mu)$ is not uniquely
characterised. The same remark applies to the formula
(\ref{rec}) for $\tilde{E}^{\rm soft}(s;\mu)$. Indeed the
inadequacy of the boundary condition (\ref{64}) to uniquely determine
the solution is already apparent in the our discussion of the formula
(\ref{pf}).

        
\appendix                               






\ack 
This work was supported by the Australian Research Council. 


\frenchspacing
\bibliographystyle{plain}

\begin{thebibliography}{99}

\bibitem{ASV95}
M.~Adler, T.~Shiota, and P.~van Moerbeke.
\newblock Random matrices, vertex operators and the {Virasoro} algebra.
\newblock {\em Phys. Lett. A}, 208:67--78, 1995.

\bibitem{AV99b}
M.~Adler and P.~van Moerbeke.
\newblock Integrals over classical groups, random permutations, {Toda} and
  {Toeplitz} lattices.
\newblock {\em Comm. Pure Appl. Math.}, 54:153--205, 2001.

\bibitem{Ad94}
V.E. Adler.
\newblock Nonlinear chains and {Painlev\'e} equations.
\newblock {\em Physica D}, 73:335--351, 1994.

\bibitem{BR99b}
J.~Baik and E.M. Rains.
\newblock The asymptotics of monotone subsequences of involutions.
\newblock {\em Duke Math. J.}, 109:205--281, 2001.

\bibitem{BF00}
T.H. Baker and P.J. Forrester.
\newblock Random walks and random fixed point free involutions.
\newblock Preprint, 2000.

\bibitem{BFP98}
T.H. Baker, P.J. Forrester, and P.A. Pearce.
\newblock Random matrix ensembles with an effective extensive external charge.
\newblock {\em J. Phys. A}, 31:6087--6101, 1998.

\bibitem{BD00}
A.~Borodin and P.~Deift.
\newblock Fredholm determinants of a class of integrable operators are
  {Jimbo-Miwa-Ueno} tau-functions.
\newblock In preparation, 2000.

\bibitem{Co00}
C.M. Cosgrove.
\newblock Chazy classes {IX-XI} of third-order differential equations.
\newblock {\em Stud. in Appl. Math.}, 104:171--228, 2000.

\bibitem{CS93}
C.M. Cosgrove and G.~Scoufis.
\newblock Painlev\'e classification of a class of differential equations of the
  second order and second degree.
\newblock {\em Stud. in Appl. Math.}, 88:25--87, 1993.

\bibitem{Fo01}
P.J. Forrester.
\newblock {\em Log-gases and {Random} {Matrices}}.
\newblock {Book} in preparation.

\bibitem{Fo93c}
P.J. Forrester.
\newblock Exact results and universal asymptotics in the {Laguerre} random
  matrix ensemble.
\newblock {\em J. Math. Phys.}, 35:2539--2551, 1994.

\bibitem{Fo93a}
P.J. Forrester.
\newblock The spectrum edge of random matrix ensembles.
\newblock {\em Nucl. Phys. B}, 402:709--728, 1993.

\bibitem{Fo95b}
P.J. Forrester.
\newblock Normalization of the wave function for the {Calogero-Sutherland}
  model with internal degrees of freedom.
\newblock {\em Int. J. Mod. Phys. B}, 9:1243--1261, 1995.

\bibitem{Fo99a}
P.J. Forrester.
\newblock Inter-relationships between gap probabilities in random matrix
  theory.
\newblock Preprint, 1999.

\bibitem{Fo99b}
P.J. Forrester.
\newblock Painlev\'e transcendent evaluation of the scaled distribution of the
  smallest eigenvalue in the {Laguerre} orthogonal and symplectic ensembles.
\newblock nlin.SI/0005064, 2000.

\bibitem{FH94}
P.J. Forrester and T.D. Hughes.
\newblock Complex {Wishart} matrices and conductance in mesoscopic systems:
  exact results.
\newblock {\em J. Math. Phys.}, 35:6736--6747, 1994.

\bibitem{FW00}
P.J. Forrester and N.S. Witte.
\newblock Application of the $\tau$-function theory of {Painlev\'e} equations
  to random matrices: {PIV}, {PII} and the {GUE}.
\newblock {\em Comm. Math. Phys.}, 219:357--398, 2001.

\bibitem{FW99}
P.J. Forrester and N.S. Witte.
\newblock Exact evaluation of the spacing distribution for random matrix
  ensembles in the bulk.
\newblock {\em Lett. Math. Phys.}, 53:195--200, 2000.

\bibitem{FW01}
P.J. Forrester and N.S. Witte.
\newblock $\tau$-function evaluations of gap probabilities in orthogonal and
  symplectic matrix ensembles.
\newblock Preprint, 2001.

\bibitem{HS99}
L.~Haine and J.-P. Semengue.
\newblock The {Jacobi} polynomial ensemble and the {Painlev\'e} {VI} equation.
\newblock {\em J. Math. Phys.}, 40:2117--2134, 1999.

\bibitem{JM81}
M.~Jimbo and T.~Miwa.
\newblock Monodromony preserving deformations of linear ordinary differential
  equations with rational coefficients {II}.
\newblock {\em Physica}, 2D:407--448, 1981.

\bibitem{JMMS80}
M.~Jimbo, T.~Miwa, Y.~M\^ori, and M.~Sato.
\newblock Density matrix of an impenetrable {Bose} gas and the fifth
  {Painlev\'e} transcendent.
\newblock {\em Physica}, 1D:80--158, 1980.

\bibitem{Jo99a}
K.~Johansson.
\newblock Shape fluctuations and random matrices.
\newblock {\em Commun. Math. Phys.}, 209:437--476, 2000.

\bibitem{Jo01}
I.M. Johnstone.
\newblock On the distribution of the largest principal component.
\newblock Preprint, 2000.

\bibitem{K99}
K.~Kajiwara, T.~Masuda, M.~Noumi, Y.~Ohta, and Y.~Yamada.
\newblock Determinant formulas for the {Toda} and discrete {Toda} equations.
\newblock {\em Funkcialaj Ekvacioj}, 44:291--307, 2001.

\bibitem{NSKGR-96}
F.~Nijhoff, J.~Satsuma, K.~Kajiwara, B.~Grammaticos, and A.~Ramani.  
\newblock A study of the alternate discrete {P}ainlev{\'e} II equation.
\newblock {\em Inverse Prob.}, 12:697--716, 1996.

\bibitem{NY99}
M.~Noumi and Y.~Yamada.
\newblock Symmetries in the fourth {P}ainlev{\'e} equation and {O}kamoto polynomials.
\newblock {\em Nagoya Math. J.}, 153:53--86, 1999.

\bibitem{NY98}
M.~Noumi and Y.~Yamada.
\newblock Higher order {Painlev\'e} equations of type {$A_l^{(1)}$}.
\newblock {\em Funkcial. Ekvac}, 41:483--503, 1998.

\bibitem{Ok86}
K.~Okamoto.
\newblock Studies of the {Painlev\'e} equations. {III}. {Second} and fourth
  {Painlev\'e} equations, {$P_{II}$} and {$P_{IV}$}.
\newblock {\em Math. Ann.}, 275:221--255, 1986.

\bibitem{Ok87}
K.~Okamoto.
\newblock Studies of the {Painlev\'e} equations. {II}. {Fifth} {Painlev\'e}
  equation {$P_{V}$}.
\newblock {\em Japan J. Math.}, 13:47--76, 1987.

\bibitem{Ok87a}
K.~Okamoto.
\newblock Studies of the {Painlev\'e} equations. {IV}. {Third} {Painlev\'e}
  equation {$P_{III}$}.
\newblock {\em Funkcialaj Ekvacioj}, 30:305--332, 1987.

\bibitem{Ra98}
E.M. Rains.
\newblock Increasing subsequences and the classical groups.
\newblock {\em Elect. J. of Combinatorics}, 5:\#R12, 1998.

\bibitem{RGTT-2000}
A.~Ramani, B.~Grammaticos, T.~Tamizhmani, and K.M.~Tamizhmani.
\newblock On a transcendental equation related to {P}ainlev{\'e} III, and its
  discrete forms.
\newblock {\em J. Phys. A:Math. Gen.}, 33:579--590, 2000.

\bibitem{ROG-2000}
A.~Ramani, Y.~Ohta, and B.~Grammaticos.
\newblock Discrete integrable systems from continuous {P}ainlev{\'e} equations
  through limiting procedures.
\newblock {\em Nonlinearity}, 13:1073--1085, 2000.

\bibitem{TW94c}
C.A. Tracy and H.~Widom.
\newblock Fredholm determinants, differential equations and matrix models.
\newblock {\em Commun. Math. Phys.}, 163:33--72, 1994.

\bibitem{TW94b}
C.A. Tracy and H.~Widom.
\newblock Level-spacing distributions and the {Bessel} kernel.
\newblock {\em Commun. Math. Phys.}, 161:289--309, 1994.

\bibitem{TW96}
C.A. Tracy and H.~Widom.
\newblock On orthogonal and symplectic matrix ensembles.
\newblock {\em Commun. Math. Phys.}, 177:727--754, 1996.

\bibitem{TW99}
C.A Tracy and H.~Widom.
\newblock On the distributions of the lengths of the longest monotone 
  subsequences in random words.
\newblock {\em Probab. Theory Related Fields}, 119:350--380, 2001.

\bibitem{TW99a}
C.A. Tracy and H.~Widom.
\newblock Random unitary matrices, permutations and {Painlev\'e}.
\newblock {\em Commun. Math. Phys.}, 207:665--685, 1999.

\bibitem{Wa98}
H.~Watanabe.
\newblock Defining variety and birational canonical transformations of the
  fifth {Painlev\'e} equation.
\newblock {\em Analysis}, 18:351--357, 1998.

\bibitem{WW65}
E.T. Whittaker and G.N. Watson.
\newblock {\em A Course of Modern Analysis}.
\newblock CUP, Cambridge, 2nd edition, 1965.

\bibitem{WFC00}
N.S. Witte, P.J. Forrester, and C.M. Cosgrove.
\newblock Gap probabilities for edge intervals in finite {Gaussian} and
  {Jacobi} unitary matrix ensembles.
\newblock {\em Nonlinearity}, 13:1439--1464, 2000.

\end{thebibliography}



                                

\end{document}